\documentstyle[12pt]{article}
\begin{document}
\titlepage

\title{\begin{flushright}
Preprint PNPI-2544,  2003
\end{flushright}
\bigskip\bigskip\bigskip
On cruel mistakes in the calculation of multi-loop
superstring amplitudes, the ambiguity of the modular integral
and the integration over the
module space} \author{G.S. Danilov \thanks{E-mail address:
danilov@thd.pnpi.spb.ru}\\ Petersburg Nuclear Physics
Institute,\\ Gatchina, 188300, St.-Petersburg, Russia}

\maketitle
\begin{abstract}
Widely spread cruel misconceptions and
mistakes in the calculation of multi-loop superstring amplitudes are
exposed.  Correct calculations are given.  It is shown that the
cardinal mistake in the gauge fixing procedure presents ab ovo in the
Verlinde papers.  The mistake was reproduced in following proposals
including the recent papers. The modular symmetry of the multi-loop
superstring amplitudes is clarified, an incorrectness of previous
conjectures being shown. It is shown that the Berezin-type integral
versus boson and fermion moduli is doubt under non-split
transformations mixing fermion integration variables to the boson
integration ones. In particular, due to singularities in moduli of the
given spin structure, the integral can be finite or divergent
dependently on the integration variables employed. Hence, unlike naive
expectations, the multi-loop superstring amplitude is ambiguous.
Nevertheless, the ambiguity is totally resolved by the requirement to
preserve local symmetries of the superstring amplitude.  In the
Verlinde world-sheet description it includes, among other thing,
the requirement that the amplitude is independent of the gravitino
field locations.  In action the resolution of the ambiguity in the
Verlinde scheme is achieved by going to the supercovariant gauge.
As it has been argued earlier, the resulted arbitrary-loop amplitudes
are finite.

\end{abstract}

\newpage

\section{Introduction}

The calculation of the multi-loop
superstring amplitudes is mainly complicated by the presence
of Grassmann moduli. In this case the amplitude is given by a
Berezin-type integral, which is doubt under non-split
transformations mixing fermion variables to the boson
ones. On the other side, non-split transformations of integration
variables are needed to provide the supersymmetry. The doubt of the
above integral under non-split transformations just is the main
difficulty for the multi-loop superstring calculations.

In particular, Ramond-Neveu-Schwarz multi-loop amplitudes each are
represented by the integral versus boson and Grassmann moduli
\cite{bshw,fried} as well as versus the interaction vertex coordinates
on a supermanifold. The integrand is the local amplitude given by a sum
over spin structures.  The local amplitude is calculated for the
reference zweibein and gravitino fields, but owing to the world-sheet
local symmetries, the whole amplitude is bound to be independent of the
reference fields above. Moreover, the local symmetry group of the
(super)string is large enough to calculate the local (super)string
amplitude (apart from a constant factor in every spin structure)
directly from the requirement that the whole amplitude is independent
with respect to local variations of the reference fields \cite{dan90}.
In this way the local amplitudes were obtained \cite{danphr,dan96} for
all spin structures given by super-Schottky groups \cite{vec,dan93} on
the $(1|1)$ complex supermanifold \cite{bshw}. The constant factors
where determined from a factorization when the handles are moved away
from each other (really it is the factorization in the poles due to
one-particle intermediate states). The calculation of the integral in
the discussed supercovariant description \cite{bshw,vec,dan93}
has been considered, too \cite{danpr1}. In doing so the
integral over the fundamental region has been added by boundary terms,
which are necessary to provide the world-sheet local supersymmetry.

It is of crucial importance that due to singularities in moduli for the
particular spin structure, the above integral can be divergent or finite
depending on the employed integration variables when they are related
to each other by a non-split transformation.  Berezin integrals of the
discussed type are avoided by mathematicians for a study.  Indeed, they
consider solely the case when the integrand is finite with all its
derivatives (in addition, they assume it to be nullified on the
boundary of the integration region that also is not the case
discussed).  Moreover, except the multi-loop superstring
calculations, the non-split changes of the integration variables for
the singular integral have not been apparently  meet in the physics.
Hence it is a new object in the physics and in the mathematics, as
well.

So, contrary to the common believer, the higher loop superstring
amplitude is ambiguous. The ambiguity is, however, totally resolved by
the requirement to preserve the local symmetries of the amplitude. In
the supercovariant description one obtains the finite and reasonable
amplitude integrating step-by step versus the super Schottky group
variables assigned to every handle \cite{danpr1}.

Now we discuss the very popular scheme where the string world-sheet is
mapped on the Riemann surface endowing with a spin structure, Grassmann
moduli being carried by the gravitino field.  The previous calculations
ab ovo \cite{ver,as} are based on unreasonable conjectures blowing by
the superficial analogy with the boson string. This analogy has created
the totally mistakable concepts on the quantum superstring, which are
widely speared by now. Therefore we explain the mistakes and give
correct calculations.

In the considered non-supercovariant description \cite{ver} the
supersymmetry is hidden, and at first sight the world sheet resembles
the world sheet of the boson string. Unexpectedly, in this case the
amplitude has been obtained \cite{ver,as,momor} depending on the
gravitino field locations. An extension of \cite{ver} to the
Green-Schwarz superstring \cite{ber} has been too unsuccessful,
in particular, because of non-physical singularities arising in
the integration measure. In fact we shall see that all the above
calculations are incorrect because of a fundamental mistake made from
the outset \cite{ver} in those gauge fixing terms, which are due to
ghost zero modes (they treat the above terms as "super Beltrami
differentials").  Nevertheless, it is true that in this world-sheet
description the interaction local amplitude is divergent and dependent
on the gravitino field locations. It is a manifestation of the
ambiguity of the Berezin integral versus moduli that was not perceived
in the discussed scheme. The calculation of the integral needs a
subtraction procedure that in action reduce the integral to the one
arising in the supercovariant scheme. With the module slice in
\cite{hok} the local vacuum amplitude  really has the
vanishing GSO projection, but in \cite{hok} the amplitude is incorrect
because there, like earlier papers, the
mistakable super Beltrami differentials were employed. In any case the
interaction amplitude can not be represented by an integral versus
boson moduli and interaction vertex coordinates on the Riemann surface
where the integrand is covariant under modular transformations.  Their
concept of the construction of the multi-loop amplitudes is an fantasy
originated by incorrect assumptions.

In particular, the mistake in super Beltrami differentials is
the result of an incorrect assumption that any expression
having a non-degenerated projection into zero modes, is acceptable for
super Beltrami differentials, as it is true for Beltrami differentials
in the boson string theory. We show that in fact super Beltrami
differentials satisfy additional restrictions, if the gravitino field
presents.

For doing so we,
as it is usual for gauge theories, calculate the jacobian of the
transformation from the integration versus the zweibein and gravitino
fields to the integration versus local gauge functions and
moduli\footnote{It is very like to the gauge fixing in QCD with the
only exception that now the global variables (moduli) present along
with gauge functions.}. The jacobian includes derivatives versus moduli
that originates the desired super Beltrami differentials. The above
jacobian was not calculated in \cite{ver}, nor in following proposals
on the matter including the
recent stream of the works \cite{hok}. Instead of that they
are guided by hazy reminiscences.
Their "super Beltrami differentials" do not satisfy necessary
restrictions, the amplitude being incorrect.

Really within many years the discussed calculations were out of a
criticism of researchers experienced in gauge theories. Perhaps, the
researches do not realize that, in particular,  multi-loop calculations
are essentially  able to correct popular nowadays conjectures based on
the tree and 1-loop calculations\footnote{The importance of the
multi-loop calculations in string theories can be demonstrated for the
non-critical boson string, as an example. In this case the introduction
of the 2D gravity makes the theory to be self-consistent up to the
1-loop approximation. In particular, the 2D gravity restores the
modular symmetry of the 1-loop integration measure. The calculation
\cite{dangr} shows, however, that the modular symmetry is remained to
be loosed for the higher-loop amplitude. So, unlike a wide opinion, the
introduction of the 2D gravity does not rescue the situation beyond the
1-loop approximation, the theory being self-discrepant.}.

An illusion of a credibility of \cite{hok} is created by the choice
of the boson moduli taken to be the elements of the period matrix on the
supermanifold in the presence of the gravitino field. As usually, the
gravitino field was given by the sum of terms, each being proportional
to the Grassmann module. So far as this module slice is invariant under
particular supersymmetry transformations containing no external
Grassmann parameters besides the Grassmann moduli, the local vacuum
amplitude is independent of the gravitino field locations. In this case
the discussed local amplitude is necessary factorized when the handles
are moved away from each other. In addition, it
is a modular form because the modular group is evidently split for
the module slice considered. The last two conditions are
sufficient to provide vanishing GSO projection  for the vacuum
amplitude in spite of the incorrect gauge fixing procedure and other
incongruous constructions of \cite{hok}. At the same time, since
Grassmann numbers are incomparable
to each other, the independence of locations of the particular  class
gravitino field does not yet ensure the independence of the general
form reference fields containing, in addition to Grassmann moduli, an
infinite number of external Grassmann parameters (being no moduli,
these external parameters are not integrated). The true result is
provided only with correct super Beltrami differentials given in the
present paper. Moreover, the incorrect
gauge fixing procedure being used \cite{hok},
the 4-point amplitude has no reasons to be reasonable, but they do
not calculated it (with an exception of its trivial piece).

Using the same module slice, we derive the correct 2-loop vacuum
amplitude and the 2-loop interaction amplitude, as well. The vacuum
amplitude is different from that claimed in \cite{hok}, but it again is
the modular form with the vanishing GSO projection. And it is
independent of the gravitino field locations. The dependence
on the gravitino field locations for the interaction amplitude
disappears only once the integration versus the interaction vertex
coordinates has been performed.

The discussed module slice exists only on the genus-2 and genus-3
surfaces where the period matrix is one-to-one with boson moduli. In
the general case the amplitude can not be represented by the integral
versus boson moduli of the modular covariant expression, which,
presumably, is independent of the gravitino field location.
The idea of such a representation lives only due to the confused
concepts \cite{as}.
Really
they \cite{as,hok} have no idea of the modular transformations
of the multi-loop superstring amplitude.

Our consideration based on the Ward relations for the local amplitude,
which are derived from the requirement that the whole amplitude is
independent of local variations of the reference fields. Like
\cite{danphr}, the local amplitude can be calculated directly by the
Ward relations discussed. The module independent multiplier in every
spin structure is determined at zero Grassmann moduli from the
condition that in this case the discussed local amplitude is modular
covariant.  In this way it is established the correspondence between
the supercovariant gauge \cite{vec,danphr,danpr1} and the
non-supercovariant parameterization \cite{ver,hok} where Grassmann
moduli are carried by the gravitino field.  Until now the desired
correspondence was absent due to the mistake for "super Beltrami
differentials" in \cite{ver,hok}.

The true super Beltrami differentials being used, the correlator for
integer-spin ghost fields differs from the correlator in \cite{ver}.
Unlike the correlator in \cite{ver}, it does not depends on locations
of Beltrami differentials, and, so, it has no non-physical poles.
Instead of that it has discontinuities in the coordinate of the
$(-1)$-ghost field. The discussed correlator is easy given in terms of
Schottky parameters \cite{dannph}.  It is related with the correlator
\cite{ver} in a non-local way including integrals along
non-contractible cycles. The reverse relation is the local one. The
correlator of the half-integer spin ghost fields is the same as in
\cite{ver}. It can be locally expressed in terms of another correlator,
which being independent of the gravitino field locations, has no
non-physical poles. Instead of that it has discontinuities in the
coordinate of the $(-1/2)$-ghost field.  Like the kindred correlator
for the integer spin fields, it is easy given in terms of Schottky
parameters \cite{dannph}. The construction of the
correlators through Schottky parameters does not use any bosonization
procedure. The fundamental domain over the Schottky
moduli is discussed in the present paper. So the local amplitude is
naturally obtained through Schottky variables.

Moreover, for the actual calculation of the interaction amplitude the
1-differentials and period matrix need to be related with the
transition group parameters as far as they determine the integration
region for the interaction vertex coordinates. In the general case the
desired expressions can be obtained \cite{vec,danphr} only through
Schottky group variables. This point was passed over in silence in
\cite{ver} and in all following to \cite{ver} proposals. So it is not
quite clear the manner in which they have intended to calculate the
more than 2-loop amplitudes. In practice, solely a Schottky-like
parameterization is useful for the calculation of the multi-loop
interaction amplitude.
Nevertheless, to be compared with the
above papers, we establish a correspondence between the result in this
parameterization and the the result in terms of the
theta-like functions \cite{ver}, which they employed.  For the actual
calculation the 1-differentials and the period matrix must both be
expressed \cite{vec,danphr} through Schottky variables.

Hence the theta-function representation is awkward in a practical
calculation of the amplitudes.  The complex variable
theory is convenient instead of that. We used it in the supercovariant
calculations \cite{danphr,danpr1,dannph}, and we use it in the present
paper, as well. Already over many years  the  complex variable theory
is usual for the calculations in the particle physics, but, perhaps, it
has passed by the superstring people operating solely by expressions
from theta-function handbooks.

In fact the simplicity of the discussed non-covariant description is
imagined. Really in this case the integrand (at least for more than 3
loops) obtained  having divergences in moduli, and, in addition, it
depends on the gravitino field locations. The actual way to remove
both these troubles is the going to the supercovariant gauge
\cite{vec,dan93} by the change of the
integration variables. Then the relevant integration procedure
\cite{danpr1} is used\footnote{
Conceptually, in the non-supercovariant
word-sheet description \cite{ver} the divergences could be removed by
the subtraction procedure preserving local symmetries of the amplitude.
Then the dependence on the gravitino field location disappears, too. It
is rather reminiscent QED and QCD where the gauge boson mass fixed to
be zero by the gauge symmetry. But the local symmetry group of the
superstring is large enough to fix all the subtraction constants. In
addition, the divergences can be avoided by a choice of the integration
variables, but in practice it is a too hard way for the more than
3-loop amplitude.}.

Important properties of local
amplitude in the non-covariant description are clarified when it is
represented through the corresponding
expressions in the supercovariant gauge, as it is performed in the
present paper.
Generally,
the present paper does not requires to use totally explicit formulas
in the above supercovariant gauge, but we would like to stress that
these expressions were obtained in \cite{danphr,danpr1,dannph}.
Especially, we recommend Appendices in \cite{danpr1,dannph} where the
integration measure and  vacuum correlators were collected.

In Sec.2 the correct fixing of the gauge of the zweibein and gravitino
fields is given. By above, we take in mind a Schottky-like
parameterization for the Riemann surface. Nevertheless, to discuss
different module slices we consider moduli to be holomorphic non-split
functions of Schottky moduli and of Grassmann ones, as well.

In Sec. 3 the restriction for the super Beltrami differentials is
obtained.

In Sec. 4 the equation for the local amplitude is derived, and the
relation between two discussed scheme is established. In doing so we
solve the task of restoring of a holomorphic function on the
non-trivial surface  from its discontinuities. In more details
tasks of this sort are discussed in the Appendix D of
\cite{danphr} and in the paper \cite{dannph}.

In Sec. 5 the local amplitude is given using
the expressions in \cite{danphr,danpr1,dannph} and in
\cite{ver}.  We give detailed references on the above expressions where
it is necessary.

In Sec. 6 the integration region and the boundary terms for the module
integral are obtained, and the modular transformations for the
superstring amplitude are discussed.

In Sec. 7, there is discussed the calculation of the 2-loop amplitude
for the moduli slice used in \cite{hok} and for other module slices
kindred to it where the local amplitude is the modular covariant. In
particular, we give the correct expression for the vacuum amplitude
instead of the mistakable one in \cite{hok}.

In Sec. 8 the ambiguity of the integral is
demonstrated, and the resolving of the ambiguity is
discussed.

Details are given in Appendices. In particular, Appendix C includes
correct expression for the 2-loop vacuum amplitude in the non-covariant
description \cite{ver}.

\section{Fixing of the gauge}

We start with the amplitude given by the integral \cite{pol} versus
both zweibein  and the world-sheet gravitino field, and versus the
string fields, as well. There are no additional integrations versus
module parameters since the zweibein and the world-sheet gravitino
field both are locally arbitrary. In doing so we map
\cite{dan90,danphr} the Riemann surface onto the complex plane $w$
choosing the transition group $w\to K(w)$ to be the same for all
surfaces of the given genus-$n$. For simplicity we can take the
transition group to be split. The integral is divided by the volume of
the local symmetry group ${\cal G}$. The group consists of world-sheet
re-parameterizations, world-sheet local SUSY  transformations, local
Lorentz transformations in the tangent space, the Weyl transformations
and $\gamma_m\iota$-shifts of the gravitino field where $\gamma_m$ is
the world-sheet Dirac matrix and $\iota$ is world-sheet spinor.  The
zweibein\footnote{Through the paper the overline denotes the complex
conjugation.} being $\tilde e_m^a(w,\overline w)$, and the gravitino
field $\tilde\phi_m(w,\overline w)$ are both represented through  the
given reference fields and the $\{\Phi\}$ set of gauge functions. Then
we go from the integration versus both  $\tilde e_m^a(w,\overline w)$
and $\tilde\phi_m(w,\overline w)$ to the integration over the gauge
functions and over moduli. At this step we employ globally defined
${\cal G}$-transformations. In this case the transition group on the
Riemann surface is not changed, and, therefore, is remained to be the
same for all the surfaces of the given genus $n$. Then the reference
fields $\{e_m^a(w,\overline w;\{q_P,\overline q_P\}),
\,\phi_m(w,\overline w;\{q_P,\overline q_P\})\}$ depend on
$(3n-3|2n-2)$ complex moduli $q_P$ (defined up to modular
transformations). Otherwise they are arbitrary that will be valuable
for the deriving of the Ward relations, which will be used for the
calculation the local amplitude.

The jacobian $J$ of the transformation is given by the superdeterminant
\begin{equation}
J=sdet\left(\begin{array}{c}
\delta\tilde e_m^a/\delta\Phi,\,
\partial\tilde e_m^a/\partial q_P,\,
\partial\tilde e_m^a/\partial\overline q_P \\
\delta\tilde\phi_m/\delta\Phi,
\partial\tilde\phi_m/\partial q_P,\,
\partial\tilde\phi_m/\partial\overline q_P
\end{array}
\right)
\label{jac}
\end{equation}
containing the derivatives versus the local gauge functions and
the derivatives versus the moduli, as well. The jacobian is represented
by the integral over ghost fields and global variables
$(\Lambda_P,\overline\Lambda_P)$ dual to $(q_P,\overline q_P)$.
It is useful \cite{dan90,howe} to combine the zweibein and the
gravitino field into a superbein $E_M^A(f,\overline f)\equiv
E_M^A(\{f^N\})$ where $f=(w|\hat\tau)$ and $f^N=(f,\overline
f)$.  Here $M$ (and $N$) labels world-sheet components while $A$ labels
the tangent space vector and spinor ones. So $M=(m,\mu_j)$ and
$A=(a,\alpha_j)$ where the Greece letters are assigned to the spinor
components.  Denoting the former $f^N$ as $f_1^N$, one obtains the
change of the superbein under the ${\cal G}$-group transformation given
by the $\{\Phi\}$ set of the local gauge functions, as it follows
\begin{equation}
E_M^A(\{f_1^N\})=\sum_{N_1,B}\frac{\partial f^{N_1}}{\partial
f_1^M} E_{N_1}^B(\{f^N\})U_B^A (\{f^N\},\{\Phi\})
\label{tra}
\end{equation}
where $U_B^A(\{f^N\},\{\Phi\})$ is a local matrix, and
$f_1^N\equiv f_1^N(\{f^N\},\{\Phi\})$. The
derivatives with respect to Grassmann co-ordinates are the "left"
ones. We use the superbein from \cite{dan90}
rather than that given in \cite{howe}. Being covariant under the whole
symmetry group, the superbein \cite{dan90} is more convenient for
the calculation than that in \cite{howe}. Unlike \cite{howe}, the
superbein \cite{dan90} contains no an additional scalar field
\cite{dan90}. Final results are the same for both \cite{howe} and
\cite{dan90}. Although a fully covariant superconnection does not
exist, one can construct two fully covariant differential operators
$\widetilde{D}$ and $D_b^a$, which only are needed for the superstring
theory. The first operator acts on a superscalar, and the second
operator determines  a change of the superbein under an
infinitesimal transformation from ${\cal G}$. In more details\footnote{
Like \cite{dan90}, we use matrices $\rho^0=-i\sigma_2$,
$\rho^1=\sigma_1$, $\rho^{\pm}=(\rho^1\pm\rho^0)/\sqrt2$ and
$\sigma_3$.  In this case $(\sigma_1,\sigma_2,\sigma_3)$ are the Pauli
matrices.  For any Majorana spinor $\eta$ we define $(\hat\eta)^\alpha=
\eta_\beta(\rho^0)^\alpha$.}
\begin{equation}
\widetilde{D}=\sum_ME^M\frac{\partial}{
\partial f^M}\,,\quad D_b^a=\rho_b\rho^a
\Biggl(\widetilde{D}-\frac{2}{s}
\sum_M\partial_M(sE^M)\Biggl)\,\,
{\rm and}\,\,
s=sdet(E_M^A),\,\,\partial_M=\frac{\partial}{\partial f^M}
\label{difop}
\end{equation}
where $E^M\equiv E^M(\{f^N\})$ is a spinor tangent space component of
the inverse superbein $E_A^M$ and derivatives are the "left" ones.
Once the integration over $\{\Phi\}$ being performed, the amplitude is
given by the following integral of the sum over spin structures $L$ and
$L'$ of right movers and of the left ones as\footnote{Through the paper
the sum over twice repeated indices is usually implied but over twice
repeated $(L,L')$ and over labels in function arguments. The otherwise
is either noted, or surely evident. Furthermore, $\{h_Y\}$ denotes the
set of certain $h_Y$ quantities where $Y$ runs the values in question.}
\begin{eqnarray}
A_n=\int(d^2q_P)\sum_{L,L'} F_{L,L'}'(\{q_P,
\overline q_P\};\{e_m^a,\phi_m\})\,,
\nonumber\\
F_{L,L'}'(\{q_P,\overline
q_P\};\{e_m^a,\phi_m\})=\int({\cal D}\Omega)(d^2\Lambda_P)V\exp
\widetilde S
\label{am}\\
\widetilde S= \int
s\Biggl[(\widetilde D\rho^0X^{\cal N})
\widetilde DX_{\cal N}+\hat B_a\biggl(D^a_bC^b
+\rho^a\tilde C\biggl)
\nonumber\\
+\hat B_a\rho_b\rho^a
E^M(\{f^N\};\{q_P,
\overline q_P\})\frac{\partial E_M^b (\{f^N\};
\{q_P,
\overline q_P\})}{\partial q_{R\ell}}\Lambda_{R\ell}
\Biggl]d^2w\,d^2\hat\theta
\label{act}
\end{eqnarray}
where $\ell=\pm$, $q_{R+}=q_R$, $\Lambda_{R+}=\Lambda_R$,
$q_{R-}=\overline q_R$ and
$\Lambda_{R-}=\overline\Lambda_R$.
In this case
$({\cal D}\Omega)$ is the product of
the string and ghost field differentials  multiplied by
the field volume form. In
addition, the integration is performed over
$(\Lambda_M,\overline\Lambda_M)$ global parameters each
to be associated with the corresponding module.
Ghosts $\hat B_a$ and $C^a$ are $(3/2)$- and, respectively,
$(-1)$-superfields while
$E_A^M(\{f^N\};\{q_P,
\overline q_P\})=(E_a^M(\{f^N\};\{q_P,
\overline q_P\}),
E^M(\{f^N\};\{q_P,
\overline q_P))$ is the inverse superbein.
Moreover,
$V$ is the
interaction vertex product integrated over the supermanifold,
and $X_{\cal N}$ is the scalar superfield where ${\cal N}$ runs from 0
till 9.
Other definitions are in
(\ref{difop}) and in footnote 5.  The derivative in respect to
Grassmann moduli is the "right" one.
The integration over $\tilde C$ gives
$\hat B_a\rho^a=0$. The last term inside the square brackets in
(\ref{act}) resembles super-Beltrami differentials, but the
zweibein in (\ref{act}) has not the conformal plan form, and
the transition group transformations $f^M\to K_s(\{f^N\})$
each assigned to the corresponding non-contractable cycle $s$, are
independent of $q_P$.  Hence $q_P$ appears only through the references
fields. The going to the conformal flat zweibein is achieved by a
suitable $f^M\to t^M$ transformation where $f^M$ depends on $q_m$ and
may depend on Grassmann moduli, as well. The local anomaly no to be,
only the last term in the square brackets in (\ref{act}) is changed due
to derivatives with respect to the moduli. Then (\ref{am})
is turned out to be
\begin{eqnarray}
A_n=\int(d^2q_P)\sum_{L,L'}{\cal F}_{L,L'}(\{q_P,
\overline q_P\};\{\phi_m\})\,,
\nonumber\\
{\cal F}_{L,L'}(\{q_P,\overline
q_P\};\{\phi_m\})=\int({\cal D}\Omega)(d^2\Lambda_P)V\exp
\widetilde S\,,
\label{a1m}\\
\widetilde{S}= \int
s\Biggl[(\widetilde D\rho^0X^{\cal N})
\widetilde DX_{\cal N}+\hat B_a\biggl[D^a_bC^b
-D^a_bv_{P\ell}^b(\{t^N\};\{q_R,
\overline q_R\})\Lambda_{P\ell}
+\rho^a\tilde C\biggl]
\nonumber\\
+\hat B_a\rho_b\rho^a
E^M(\{f^N\};\{q_R,
\overline q_R\})\frac{\partial E_M^b (\{f^N\};
\{q_R,
\overline q_R\})}{\partial q_{P\ell}}\Lambda_{P\ell}
\Biggl]d^2w\,d^2\hat\theta \,,
\label{ac1t}\\
v_{P\ell}^b(\{t^N\};\{q_R,
\overline q_R\})=
v_{P\ell}^M(\{t^N\};\{q_R,
\overline q_R\})E_M^b(\{t^N\})\,,
\nonumber\\
E_m^b=e_m^b+\hat\theta\rho^b\phi_m\,,\quad
E_\alpha^b=-(\rho^b\theta)_\alpha\,,\quad e_m^b
=e\delta_m^b
\label{vecb}
\end{eqnarray}
where \cite{dan90} $E_M^b\equiv E_M^b(\{t^N\})$,
$D^a_b$ is the same as in (\ref{difop}), and
\begin{equation}
v_{P\ell}^M(\{t^N\};
\{q_R,
\overline q_R\})=\sum_{J}\tilde\epsilon(PJ)\frac{\partial
f^J(\{t^N\};
\{q_R,
\overline q_R\})}{\partial q_{P\ell}} \frac{\partial
t^M}{\partial f^J}
\label{vect}
\end{equation}
where
$\tilde\epsilon(RP)=1$, if $R$ and $P$ both label a boson
(fermion). Otherwise $\tilde\epsilon(RP)=-1$.
In the
calculation one uses that
\begin{equation}
\sum_{M,N}\Biggl[
\tilde\epsilon(M(N+R))E^Mv_{R\ell}^N\partial_ME_N^a-
\tilde\epsilon(MR)
E^Mv_{R\ell}^N\partial_NE_M^a\Biggl]
-\sum_M\frac{2}{s}(\partial_MsE^M)
v_{R\ell}^a=0
\label{tens}
\end{equation}
where $v_{R\ell}^N\equiv v_{P\ell}^N(\{t^M\};
\{q_P,
\overline q_P\})$ and
$\tilde\epsilon(RP)$  is the same  as in (\ref{tens}).
It is sufficient to check (\ref{tens}) for
$e_m^a=\delta_m^a$ and $\phi_m=0$ since its right side
is covariant under local transformations of the ${\cal G}$
group (if the superbein \cite{dan90} is used).
The term with $v_{R\ell}^b(\{t^M\};
\{q_P,
\overline q_P\})$ in (\ref{ac1t})
is originated by the above discussed addition to the last term in
(\ref{act}). Once the integration over $\tilde C$ has been performed,
the $\hat B_a\rho^a=0$ restriction arises. Furthermore, if
$t^M\to\tilde t^M(\{t^N\})$ and $q_{R\ell}\to\tilde
q_{R\ell}(\{q_R,\overline q_R\})$ then
$v_{P\ell}^M(\{t^N\};
\{q_R,
\overline q_R\})\to \tilde v_{P\ell}^M(\{\tilde t^N\};
\{\tilde q_R,
\overline{\tilde q_R}\})$ where, from (\ref{vect}),
\begin{equation}
v_{P\ell}^M(\{t^N\};
\{q_Q,
\overline q_Q\})=
\biggl[\tilde\epsilon(M(R+P))\tilde v_{R\ell'}^N(\{t^J\};
\{\tilde q_Q,
\overline{\tilde q_Q}\})\frac{\partial
t^M}{\partial\tilde  t^N}
-\tilde\epsilon(MP)\frac{\partial t^M}{\partial
\tilde q_{R\ell'}}\biggl]\frac{\partial
\tilde q_{R\ell'}}{\partial q_{P\ell}}
\label{trvec}
\end{equation}
Unlike
the transition group transformations $f^M\to K_s^M(\{f^N\})$ on the $f$
supermanifold, the transition group ones $t^M\to\Gamma_s^M(\{t^N\})$
depend on the moduli. Thus $v^b$ has discontinuities on the
$t$-supermanifold. Indeed, from (\ref{trvec}) it follows that
\begin{equation}
v_{P\ell}^M(\{\Gamma_s^N\};
\{q_R,
\overline q_R\})=v_{P\ell}^J(\{t^N\};
\{q_R,
\overline q_R\})\frac{\partial\Gamma_s^M}{\partial t^J}-
\tilde\epsilon(MP)\frac{\partial\Gamma_s^M}{\partial q_{P\ell}}
\label{vct}
\end{equation}
where $\Gamma_s^N\equiv\Gamma_s^M(\{t^N\})$. In this case one uses that
$f^M(\{\Gamma_s^N)\})=K_s^M(\{f^N(\{t^N\})\})$. The last term on the
right side of (\ref{vct}) is just the discontinuity of
$v_{P\ell}^M(\{t^N\};\{q_R,\overline q_R\})$.
The desired discontinuity is given solely by the transition group
but it is independent of the particular choice of $v_{R\ell}^b$.
As it is
usually (see also the next Section), only zero modes
of $B$ contribute to the $\sim v_{R\ell}^b$ term in (\ref{ac1t}). Thus
$\tilde S$ depends only on the discontinuities of
$v_{R\ell}^b$, and, therefore, it is independent of
the particular choice of $v_{R\ell}^b$.

In the supercovariant gauge \cite{vec,dan93} the gravitino field
vanishes (modulo $\gamma_m\iota$). In this case the transition
group transformations depend, in addition, on Grassmann moduli. Usually
spin structures are defined in terms of super-Schottky groups acting on
the complex $(1|1)$ supermanifold \cite{bshw}. Generally, a
super-Schottky transformation depends on 3 boson and 2 Grassmann
complex parameters. For non-zero Grassmann parameters fermions are
mixed to bosons providing a holomorphic world-sheet supersymmetry.

As we already discussed in the Introduction, now we mainly
consider the non-covariant scheme \cite{ver,hok} where
Grassmann moduli are carried by the gravitino field $\phi_m$.  The
gravitino field can be reduced to the given gravitino one
$\hat\phi_m^{(0)}$ by the local symmetry group changes remaining the
conformal flat zweibein, and the transition group to be independent of
Grassmann moduli. Then, generally, the boson moduli each receive a
$\phi_m$-dependent addition, and $v_{R\ell}^b$ is changed, as well. As
far as $\tilde S$ does not depend on the particular choice of
$v_{R\ell}^b$, the gravitino field $\phi_m$ is removed from (\ref{a1m})
by a relevant non-split change of the moduli provided that the boundary
integral is added, and the integration of singularities is performed
correctly. Then both descriptions are on equal terms.

We take the transverse gravitino field that is $\rho^m\phi_m^{(0)}=0$
where $\rho^m$ is defined in the footnote 5. So $\phi_m^{(0)}$ is given
by its components $\phi_{\pm}$. Global parameters and the ghosts being
relevantly re-defined, the amplitude (\ref{a1m}) in the gauge discussed
(that is Grassmann moduli are carried by the gravitino field)
is given by
\begin{eqnarray}
A_n=\int(d^2q_P')\sum_{L,L'}\widehat F_{L,L'}(\{q_R',
\overline q_R'\};\{\phi\})\,,
\nonumber\\
\widehat F_{L,L'}(\{q_R',\overline
q_R'\};\{\phi\})=\int({\cal
D}\Omega)(d^2\Lambda_P)V\exp S_\phi\,,
\label{ampl}
\end{eqnarray}
where $q_R'$ are complex moduli, and other definitions are given in
(\ref{a1m}). Furthermore\footnote{With
the re-scaling $2^{1/4}i\theta_+\to\vartheta$, $2^{1/4}\theta_-
\to\overline\vartheta$, $2^{1/4}i\phi_-\to\phi_-$,
$2^{1/4}\phi_+\to\phi_+$.}
\begin{eqnarray}
S_\phi=\int
d^2t\biggl[B(t,\bar t)\Upsilon_{P\ell}^{(+)}(t,\bar t;\{q_R, \overline
q_R\}) -\overline B(t,\bar t) \Upsilon_{P\ell}^{(-)}(t,\bar t;\{q_R,
\overline q_R\})\biggl]\Lambda_{P\ell} \nonumber\\ +\int d^2t\biggl[
B\widehat{D}_+^{(\phi)}C
-\overline
B\widehat{D}_-^{(\phi)}
\overline C\biggl]
+\int d^2tD_+^{(\phi)}X_{\cal N}D_-^{(\phi)}
X^{\cal N}
\label{actio}
\end{eqnarray}
where $B$  and $C$ are the ghost fields form $(3/2)$-superfield and,
respectively, $(-1)$-one.  As in (\ref{a1m}), we  use the notations
$q_{P\ell}$ and $\Lambda_{P\ell}$ for $(q_P,\Lambda_P)$ and their
complex conjugated.  So called super Beltrami differentials
$\Upsilon_{P\ell}^{(\pm)}(t,\bar t;
\{q_R, \overline q_R\})$ are calculated
from the corresponding terms in (\ref{ac1t}). They are discussed in the
next Section.  Moreover,
\begin{eqnarray}
D_-^{(\phi)}=D+\frac{1}{2}\phi_+\biggl[\vartheta\frac{\partial}{
\partial\overline\vartheta}+
\overline\vartheta\vartheta\frac{\partial}{
\partial\overline z}\biggl]\,,\quad
D_+^{(\phi)}=\overline{D(t)}
+\frac{1}{2}\phi_-\biggl[\overline\vartheta\frac{\partial}{
\partial\vartheta}-
\overline\vartheta\vartheta\frac{\partial}{
\partial z}\biggl]\,,
\nonumber\\
\widehat{D}_-^{(\phi)}=
D_-^{(\phi)}
-(\overline{\partial_z}\phi_+)
\overline\vartheta\vartheta\,,\quad
\widehat{D}_+^{(\phi)}=
D_+^{(\phi)}
+(\partial_z\phi_-)
\overline\vartheta\vartheta\,;\quad
D(t)=\vartheta\partial_z+\partial_\vartheta\,.
\label{derphi}
\end{eqnarray}
As
usually, we take
\begin{equation}
\phi_{\pm}=\sum_{s=1}^{2n-2}\lambda_s^{(\pm)}\phi_{s\pm}\,,
\quad \phi_{s-}=\pi\delta^2(z-z_{(s)})\,,\,\,
\phi_{s+}=\pi\delta^2(z-\overline z_{(s)}')\,,\quad
\lambda_s^{(-)}\equiv
\lambda_s,\,\,\lambda_s^{(+)}\equiv\overline
\lambda_s
\label{grfield}
\end{equation}
where $\{\lambda_s^{(m)}\}$ is the set of Grassmann moduli.
The $(z_{(s)},\overline z_{(s)}')$ locations all are different from
each other.  Generally, they depend on boson moduli.  The
$\delta$-functions may be "spread", if it is necessary to remove
an uncertainty due to the $\delta$-function localization.

To the round of the given non-contractable
cycle $s$, the transition group transformation $t\to\Gamma_s(t)$ is
assigned to be
\begin{equation}
t\to\Gamma_s(t):\quad z\to
g_s(z;\{q_r\})\,,\quad \vartheta\to\pm
\sqrt{\partial_zg_s(z;\{q_r\})} \vartheta.
\label{trgrtr}
\end{equation}
For the sake of simplicity we consider $g_s(z;\{q_r\})$ depending
on complex moduli $\{q_r\}$ but not on the complex conjugated ones.
Generally, $q_R$ and $q_R'$ can related to each other by a
superholomorphic transformation as $q_P=q_P(\{q_R'\})$.

\section{Restrictions for super Beltrami differentials}

Generally, $\widehat F_{L,L'}(\{q_R',\overline
q_R'\};\{\phi\})$ in (\ref{ampl}) can depend on $\phi_m$, but the
amplitude $A_n$ is expected to be $\phi_m$-independent. In particular,
it is provided by the super Beltrami differentials
$\Upsilon_{P\ell}^{(\pm)}(t,\bar t;\{q_R,\overline q_R\})$.
in (\ref{actio}).  Thus the super Beltrami
differentials for two sets of moduli $\{q_P\}$ and $\{q_P'\}$ with
$q_P'=q_P'(\{q_{P\ell}\})$ are related to each other by
\begin{equation}
\Upsilon_{P\ell}^{(\pm)}(t,\bar
t;\{q_R',\overline q_R'\})= \Upsilon_{Q\ell'}^{(\pm)}(t,\bar
t;\{q_R,\overline q_R\}) \frac{\partial q_{Q\ell'}} {\partial
q_{P\ell}'}\,.
\label{modmod}
\end{equation}
Furthermore, one can see that the super Beltrami differentials are
represented as
\begin{eqnarray}
\Upsilon_{P\ell}^{(+)}(t,\bar t;\{q_R, \overline q_R\})=
\overline\vartheta\vartheta
\frac{\partial\phi_-}
{\partial q_{P\ell}}- r_{P\ell}^{(+)}(t,\bar
t;\{q_R, \overline q_R\})\,,
\nonumber\\
\Upsilon_{P\ell}^{(-)}(t,\bar
t;\{q_R, \overline q_R\})= \vartheta\overline\vartheta
\frac{\partial\phi_+}
{\partial q_{P\ell}'}- r_{P\ell}^{(-)}(t,\bar
t;\{q_R, \overline q_R\})
\label{beldif}
\end{eqnarray}
where $\{q_R\}=\{q_r,\lambda_i\}$. Furthermore,
$r_{P\ell}^{(\pm)}(t,\bar t;\{q_R, \overline q_R\})$
admits the representation
\begin{equation}
r_{P\ell}^{(+)}(t,\bar t;\{q_R, \overline q_R\})
=\widehat{D}_+^{(\phi)}
v_{P\ell}^{(+)}(t,\bar t;\{q_R, \overline q_R\})\,,
\quad
r_{P\ell}^{(-)}(t,\bar t;\{q_R, \overline q_R\})
=\widehat{D}_-^{(\phi)}
v_{P\ell}^{(-)}(t,\bar t;\{q_R, \overline q_R\})\,.
\label{berdf}
\end{equation}
In this case $v_{R\ell}^{(\pm)}(t,\bar t;\{q_R, \overline q_R\})$ is
the same as $v_{R\ell}^b(\{t^N\};\{q_R, \overline q_R\})$.
The change under (\ref{trgrtr}) of $v_{R\ell}^{\pm}(\{t^N\};\{q_R,
\overline q_R\})$ is obtained from ({\ref{vct}). If, for the sake of
simplicity, we take $z$ to be out of the gravitino field location, then
$v_{\alpha\ell}^{(\pm)} (t,\bar t;\{q_R, \overline q_R\})$ has no
discontinuities while $v_{p\ell}^{(\pm)} (t,\bar t;\{q_R, \overline
q_R\})$ has the discontinuity under transformation (\ref{trgrtr}). In
particular,
\begin{equation}
v_{p\ell}^{(+)}
(\Gamma_s(t),\overline{\Gamma_s(t)};\{q_R, \overline q_R\})=
v_{p\ell}^{(+)}(t,\overline t;\{q_R, \overline q_R\})
\frac{\partial g_s(z;\{q_r\})}{\partial z}
-\frac{\partial g_s(z;\{q_r\})}{\partial q_{p\ell}}
\label{jump}
\end{equation}
where the last term on the right side is the discontinuity of
$v_{p\ell}^{(+)}(t,\overline t;\{q_R, \overline q_R\})$. In this case
$v_{p-}^{(+)}(t,\overline t;\{q_R, \overline q_R\})$ has no the
discontinuity.  The discontinuity of $v_{p\ell}^{(-)}$ is given by the
complex conjugated expression.  In particular, from (\ref{berdf}),
\begin{equation}
\int d^2t
\widehat{\chi}_Q^{(\phi+)}(t)
r_{P+}^{(+)} (t,\bar t;\{q_R, \overline q_R\})
=\int_sdt\,\widehat{\chi}_Q^{(\phi+)}(t)
[v_{P+}^{(+)}(t,\overline t;\{q_R, \overline q_R\})]_{s}
\label{matr}
\end{equation}
where $dt=dzd\vartheta$ while
$[v_{P+}^{(+)}(t,\overline t;\{q_R, \overline q_R\})]_s$ is the
discontinuity of $v_{P+}^{(+)}(t,\overline
t;\{q_R, \overline q_R\})$ around $s$-cycle, and
$\widehat{\chi}_P^{(\phi+)}(t)$ is the zero mode of
$[\widehat{D}_{+}^{(\phi)}]^T$. The right side integrals versus $dz$
each are calculated along the relevant non-contractable cycle on
$z$-plane, and the summation over $s$ is implied. We imply that
gravitino field locations are out of the non-contractable cycles above.
The kindred relations for
$r_{P-}^{(-)} (t,\bar t;\{q_R, \overline q_R\})$ contain
zero mode of $[\widehat{D}_{-}^{(\phi)}]^T$.

Eqs. (\ref{matr}) can not be satisfied solely by a choice of the moduli
as far as the $(QP)$  matrix being formed by the integrals, is
degenerated due to (\ref{berdf}) along with the above mentioned
conditions for $[v_{P+}^{(+)}(t,\overline t;\{q_R, \overline
q_R\})]_s$.  In this case $r_{P-}^{(+)} (t,\bar t;\{q_R, \overline
q_R\})$ must be partly orthogonal to relevant ghost zero modes
$\widehat{\chi}_Q^{(\phi+)}(t)$. We defined the above zero modes
through discontinuities of the ghost superfield Green function
$\widehat G_{(\phi)}^{(+)}(t,t')$  having no additional poles in
the fundamental region on $(z,z')$-complex planes except the usual pole
at $z=z'$. So
\begin{equation}
\widehat D_+^{(\phi)}
\widehat G_{(\phi)}^{(+)}(t,t')=-
\widehat G_{(\phi)}^{(+)}(t,t')[\widehat D_+^{(\phi)}]^T
=\delta^2(z-z')\delta^2(\vartheta-\vartheta')\,.
\label{zmgreq}
\end{equation}
In this case $\widehat D_+^{(\phi)}$ acts on $(\bar t,t)$ and
$[\widehat D_+^{(\phi)}]^T$ acts on $(\bar t',t')$. At $\phi_-\equiv0$
the $\widehat{G}_{(\phi)}^{(+)}(t,t')$ Green function is reduced to
$\widehat G(t,t')$, and $\widehat{\chi}_P^{(\phi+)}(t)$ is reduced to
$\widehat{\chi}_P(t)$. Furthermore,
\begin{equation}
\widehat G(t,t')
=\widetilde G_b(z,z')\vartheta'+\vartheta
\widetilde G_f(z,z')\,,
\quad\widehat\chi_r(t')=-\vartheta'\widetilde\chi_r(z')\,,
\quad\widehat\chi_\alpha(t')=\widetilde\chi_\alpha(z')
\label{zergr}
\end{equation}
where $\widetilde G_b(z,z')$ is a conformal $2$-tensor in $z'$,
$\widetilde G_f(z,z')$ is the conformal $(3/2)$-one,
$\widetilde\chi_r(z')$ is 2-tensor zero mode
and $\widetilde\chi_\alpha(z')$ is $(3/2)$-tensor zero one.
The above Green functions and zero modes
(\ref{zergr}) have been discussed in \cite{dannph}, where they
are given\footnote{The discussed Green functions each being
multiplied by $\pi$, are equal to
$-G_{gh}^{(b)}(z,z')$ and, respectively, $G_{gh}^{(f)}(z,z')$
in \cite{dannph}
while
$\pi\widetilde\chi_r(z)$ and $\pi\widetilde\chi_\alpha(z)$
with $\alpha=\alpha_s$ are
the same as $\tilde\chi_{R_r}^{(0)}(z)$ and
$\tilde\chi_{F_s}^{(0)}(z)$) in \cite{dannph}.}
through the Schottky parameters (see eqs.(33) in
\cite{dannph}.
They can also be expressed
through ghost vacuum correlators \cite{ver}
depending on the locations points see Appendix A
in the present paper).
Both they have discontinuities in $z$ as follows
\begin{eqnarray}
\widetilde
G_b(g_s(z),z')=\frac{\partial g_s(z)}{\partial z}
\left(\widetilde
G_b(z,z')+
\sum_mP_r^s(z)(z)\widetilde\chi_r(z')\right)
\label{chbgr}\\
\widetilde
G_f(g_s(z),z')=\sqrt{\frac{\partial g_s(z)}{\partial z}}
\left(\widetilde G_f(z,z')+
\sum_j
P_{\alpha_j}^{(s)}(z)\widetilde\chi_{\alpha_j}(z')\right)\,.
\label{chfgr}
\end{eqnarray}
Here $P_{\alpha_j}^{(s)}(z)$ is a first order polynomial (see
eqs. (23)--(26) in \cite{dannph}). Moreover,
\begin{equation}
\widehat{G}_{(\phi)}^{(+)}(t,t')= \widehat{G}(t,t')
-\int d^2\tilde t\widehat{G}(t,\tilde t) [\widehat{D}_+^{(\phi)}
-\overline
D]G_{(\phi)}^{(+)}(\tilde t,t')
\label{intgr}
\end{equation}
where $\widehat D_-^{(\phi)}$ and $D\equiv D(\tilde t)$ are
defined by (\ref{derphi}). The kernel being
$\sim\phi_-$, the equation is solved
by the iteration procedure. Due to zero modes,
$\widetilde G_b(z,z')$ and $\widetilde G_f(z,z')$ have
discontinuities on $z$-plane that gives rise to the
discontinuities of
$\widehat G_{(\phi)}^{(+)}(t_1,t)$. Correspondingly,
\begin{equation}
\widehat{\chi}_R^{(\phi+)}(t)=
\int d^2t_1
\widehat{\chi}_R(t_1)\overline{D(t_1)}
\widehat{G}_{(\phi)}^{(+)}(t_1,t)
\label{zmgr}
\end{equation}
where the integral versus $z_1$ is taken over the fundamental region.
The desired restrictions can be given as
follows
\begin{equation}
\int d^2t
\widehat{\chi}_\alpha^{(\phi+)}(t)
r_{n+}^{(+)} (t,\bar t;\{q_R, \overline q_R\})=0\,,\quad
\int d^2t
\widehat{\chi}_P^{(\phi+)}(t)
r_{\alpha\ell}^{(+)} (t,\bar t;\{q_R, \overline q_R\})=0
\,,\quad d^2t=d^2zd^2\vartheta
\label{orthog}
\end{equation}
Indeed, calculating
$v_{P+}^{(+)}(t,\overline t;\{q_R,\overline q_R\})$ by (\ref{berdf})
through the Green function, one can see that the first of
(\ref{orthog}) is the condition that
the discontinuity of
$v_{P+}^{(+)}(t,\overline
t;\{q_R,\overline q_R\})$ contains no the $\sim\vartheta$ term.
The second of (\ref{orthog}) is due to
$r_{\alpha\ell}^{(+)} (t,\bar t;\{q_R,\overline q_R\})$ has no
discontinuities.
Since $r_{\alpha\ell}^{(+)} (t,\bar
t;\{q_R,\overline q_R\})$ has no discontinuities, it can be
vanishing as in \cite{ver}, but in \cite{ver} the first of
(\ref{orthog}) is not satisfied. In \cite{hok} both relations
(\ref{orthog}) are not satisfied in the required basis (\ref{zmgr}).

As usually, a
shift of $C$ removes that part of
$r_{P\ell}^+(t_1,\overline t_1)$,
which is orthogonal to
zero modes of $B$. The remained
$r_{P\ell}^{(+)}(t_1,\overline t_1)$-dependent terms in
(\ref{actio}) being integrated by parts, are reduced to the
integral along $(A,B)$-cycles, which contains only of
$[v_{P\ell}^{(+)}(t,\bar t;\{q_R,\overline q_R\})]_s$.  So
$\widehat F_{L,L'}(\{q_R,\overline q_R\};\{\phi\})$ is
independent of the particular $v_{R\ell}^{(\pm)}(t,\bar
t;\{q_R, \overline q_R\})$, and explicit
$v_{R\ell}^{(\pm)}(t,\bar t;\{q_R,\overline q_R\})$ are not necessary
for the calculation of the amplitude.

If one  derives (\ref{ampl}) from the integral over both
the zweibein and the gravitino field bypassing
(\ref{am})--(\ref{act}),
then derivatives with respect to $q_{M\ell}$ in the jacobian
arise due to the $f_1^M\to t^M$ transformation depends on
$\{q_{M\ell}\}$, the final result being the same.

\section{Equations for the local amplitude}

Really the integral (\ref{ampl}) over the fields requires a
regularization, which provides the independence of the
superstring amplitude of the local alterations of the reference
fields. The desired integral can be also calculated from the
Ward relations
derived just from the requirement $\delta_\perp
A_n=0$. Here $\delta_\perp A_n$ is the alteration of the
amplitude (\ref{am}) under infinitesimal transverse local
variations $\delta_\perp e_m^a$ and $\delta_\perp\phi_m$ of
$e_m^a$ and, respectively, of $\phi_m$. In this case
$e_a^m\delta_\perp e_m^a=e_a^m\varepsilon_{ab}\delta_\perp
e_m^b=0$, $\gamma^m\phi_m=0$,
$\varepsilon_{-+}=-\varepsilon_{+-}=1$.
A kindred relation can be derived for other superstring models
including the superbrane models, as well.
In the supercovariant scheme it was
obtained early in \cite{dan90,danphr}.
The field volume form in
(\ref{am}), generally, depends on $(e_m^a,\phi_m)$, but it is
not changed under the considered variations. So the variation
$\delta_\perp A_n$ of the amplitude (\ref{am}) is due to only by
the alteration of (\ref{act}).  Going to the conformal flat
zweibein, we obtain the desired relations  for $\widehat
B_{L,L'}(\{q_M,\overline q_M\};\{\phi\})$ in (\ref{ampl}).
Since $\delta_\perp e_m^a$ and $\delta_\perp\phi_m$ are
arbitrary, the relations are local in
$t=(z|\vartheta)$.  Furthermore,
\begin{equation}
\widehat F_{L,L'}(\{q_M,\overline
q_M\};\{\phi\})=\widehat Z_{L,L'}(\{q_M,\overline
q_M\};\{\phi\})<V>_\phi
\label{bzvac}
\end{equation}
where $<V>_\phi$ is the vacuum expectation of $V$ in
(\ref{ampl}) and $\widehat Z_{L,L'}(\{q_M,\overline
q_M\};\{\phi\})$ is the vacuum function.
For simplicity, we choose $q_m'=q_m$, see (\ref{ampl}) and
(\ref{trgrtr}). As above,
$\{q_\alpha\}=\{\lambda_i\}$.
It will be sufficient to set
$\delta_\perp e_m^a$ and
$\delta_\perp\phi_m$ in the points
different from the locations
of the gravitino field (\ref{grfield}).
The deriving of the equations is very
similar to the that given in \cite{dan90} where one can see for
details. The desired equations are found to be
\begin{eqnarray}
\widehat\chi_R^{(\phi+)}(t)\frac{\partial}{\partial q_R}
\ln\widehat Z_{L,L'}(\{\phi\})
=<T>_\phi-\frac{\partial}{\partial q_R}
\widehat\chi_R^{(\phi+)}(t)\,,
\nonumber\\
\widehat\chi_R^{(\phi+)}(t)\frac{\partial}{\partial
q_R}<V>_\phi = <TV>_\phi-<T>_\phi<V>_\phi\,;
\label{equ}\\
T=-(DX^{\cal
N})\partial_zX_{\cal N} +\frac{3}{2}B\partial_zF+(\partial_z B)F
-\frac{1}{2}(DB)(DF)\,,
\label{enchi}\\
\widehat\chi_R^{(\phi+)}(t)=-
<B(t,\overline t)\Lambda_R>_\phi\,,\quad
F(t,\bar t)=C(t,\bar t)-
v_{P\ell}^{(+)}(t,\bar t;\{q_R,\overline q_R\})
\Lambda_{P\ell}
\label{fieldf}
\end{eqnarray}
along with the kindred relations due to the left
movers. Here $\widehat\chi_R^{(\phi+)}(t)$ are given by
(\ref{zmgr}),
$\widehat Z_{L,L'}(\{\phi\})\equiv
\widehat Z_{L,L'}(\{q_R,\overline q_R\};\{\phi\})$ in
(\ref{bzvac}), and $F(t,\overline t)$ has
discontinuities calculated by (\ref{jump}).
The derivatives versus
$\{q_\alpha\}$ each are to be the right-side ones.
The string superfield correlator
$<XX>_\phi$, the superghost correlator $<CB>_\phi$ and zero modes
$<B\Lambda_{R\ell}>_\phi$ in the given gravitino field are
calculated in the known manner by adding to (\ref{actio}) of the
source term $[X\tilde{{\cal X}}+B\tilde{{\cal C}}+ \tilde{{\cal
B}}C+{\cal L}_{R\ell}\Lambda_{R\ell}+c.c]$. Here $(\tilde{{\cal
B}},\tilde{{\cal C}},{\cal L}_{R\ell})$ are sources. Terms linear
either in $C$, or in non-zero modes of $B$ are removed by
shifts of $(B,C)$ using $G_\perp^{(\phi+)}(t;t_1)$, which is the
orthogonal to $\widehat{\chi}_R^{(\phi+)}(t_1)$ part of
$G^{(\phi+)}(t;t_1)$. The source dependence are extracted as
the exponential of $[-\tilde{{\cal X}}<XX>_\phi \tilde{{\cal
X}}-\tilde{{\cal B}}<CB>_\phi\tilde{{\cal C}}- {\cal L}_{R\ell}
<\Lambda_{R\ell}B>_\phi\tilde{{\cal
C}}]$.
The scalar superfield correlator is
given by the Green function of
$D_-^{(\phi)}D_+^{(\phi)}$, see (\ref{derphi}).
In doing so the holomorphic Green
functions, superscalar functions
and their periods are constructed for the
$D_+^{(\phi)}$ operator, see Appendix B for more details.
Using $<CB>_\phi$
and $<B\Lambda_{R\ell}>_\phi$, one calculates $<FB>_\phi$ in
(\ref{fieldf}). Unlike $<CB>_\phi$, both $<FB>_\phi$ and
$<B\Lambda_R>$ are independent of the particular
$v_{R\ell}^{(\pm)}(t,\bar t;\{q_R,\overline q_R\})$.
To evaluate an indetermination arising
due to the poles at the gravitino field
locations,
the integrals over the Riemann surface are previously
calculated for a "spread" gravitino field.
At $\{\lambda_i=0\}$,
\begin{equation}
-<F(t)B(t')>_{\phi=0}\equiv G(t,t')=
-\widetilde G_b(z,z')\vartheta'+
\vartheta G_{3/2}(z',z;\{z_{(i)}\})
\label{corfb}
\end{equation}
where $\widetilde G_b(z,z')$ is discussed in the previous Section
and $G_{3/2}(z',z;\{z_{(i)}\})$ is the $(3/2,1/2)$-field correlator
\cite{ver} depending on
the $\{z_{(i)}\}$ locations of $\phi_-$, see Appendix A for more
details. As it is usually, calculating $T$ in (\ref{enchi}), one omits
the singularity at the same points. Due to the $F$-superfield, $T$ has
a discontinuity under the transition group transformation.
The discontinuity is, however, canceled by the discontinuity of
$-\partial_{q_R} \hat\chi_R^{(\phi+)}(t)$, details being omitted here.
As the result, the right side of the first equation among
(\ref{equ}) has no discontinuities on $z$-plane.

Eqs. (\ref{equ}) occur as in \cite{dan90,danphr}, but in
\cite{dan90,danphr} the vacuum correlators and zero modes are
calculated on the non-split supermanifold
\cite{bshw} given by super-Schottky group
transformations $\tilde t\to\tilde\Gamma_s(\tilde t)$.  We map
the above supermanifold by $\tilde t=(\tilde
z|\tilde\vartheta)$.  The $(L,L')$ spin structures are defined
for the super-Schottky groups. To every handle one assigns
multiplier $\tilde k_s$ along with two group limiting points
$(\tilde u_s|\mu_s)$  and $(\tilde v_s|\nu_s)$ on the
supermanifold. In this case $(3|2)$ of
$(\tilde u_s,\tilde v_s|\mu_s,\nu_s)$
and of the interaction vertex coordinates on the $\tilde t$
supermanifold
are fixed due to $SL(2)$ symmetry. Below for simplicity
we fix $(3|2)$ of
$(\tilde u_s,\tilde v_s|\mu_s,\nu_s)$,
the set of $(3n-3|2n-2)$
complex moduli being $\{\tilde q_M\}$.
The amplitude is given like (\ref{ampl}) as
\begin{eqnarray}
A_n=\int(d^2q_P)\sum_{L,L'}\widetilde F_{L,L'}(\{q_R,
\overline q_R\})\,,
\nonumber\\
\widetilde F_{L,L'}(\{q_R,
\overline q_R\})=\widetilde{Z}_{L,L'} (\{\tilde
q_R,\overline{\tilde q_R}\})<V>\,,\quad
\widetilde{Z}_{L,L'}(\{\tilde q_R,\overline{\tilde q_R}\})=
\nonumber\\
={\det}^{-5}[\omega(\{\tilde q_R\};L)+ \overline{\omega(\{\tilde
q_R\};L')}] Z_L(\{\tilde q_R\})\overline{Z_{L'}(\{\tilde q_R\})}
\label{amplit}
\end{eqnarray}
where $<V>$ is the vacuum expectation of the interaction vertex
product integrated over the supermanifold,
and
$\widetilde Z_{L,L'}(\{\tilde q_R,
\overline{\tilde q_R}\})$ is the
vacuum function,
$Z_L(\{\tilde q_R\})$ being holomorphic in $q_R$.
For all the
even spin structures the vacuum functions and vacuum
correlators calculating $<V>$, both are given in
\cite{danphr,danpr1,dannph}. In this case
$\omega(\{\tilde q_R\};L)/(2\pi i)$
is the period matrix on the supermanifold.
Due to fermion-boson mixing, $\omega(\{\tilde q_R\};L)$
depends on Grassmann moduli and
on $L$.
So the holomorphic structure of the vacuum function is
straightforward only until the  integration
versus Grassmann moduli to be
performed.  The local amplitude admits the representation
by the integral with respect to loop 10-momenta, the integrated
expression being the product of a holomorphic function of the boson and
Grassmann moduli and of the interaction vertex coordinates on the
supermanifold times by the anti-holomorphic function of the
variables considered.  Transition to the $t$-supermanifold is achieved
by the $\tilde t\to t$ local symmetry group transformation remaining
the zweibein to be
conformal flat. In doing so $\tilde q_R\to q_R$.
As  far as the resulted transition group
transformations $t\to\Gamma_s(t)$ are independent of Grassmann moduli,
a gravitino field arises. Since the zweibein is conformal flat,
$\tilde t\to t$ is a super-conformal transformation out
of the gravitino field locations. As an example of the $\tilde t\to t$
transformation, we take $\tilde t(t)$ to be holomorphic in $t$ except
poles at $z=z_{(i)}$ giving rise to the gravitino field in
(\ref{grfield}). In this case \cite{bshw,dannph},
\begin{eqnarray}
\tilde
z=z+f(z)+[1+f'(z)]\vartheta\xi(z)\,,\quad
\tilde\vartheta=\sqrt{1+f'(z)}\biggl[(1+\frac{1}{2}\xi(z)\xi'(z))
\vartheta+\xi(z)\biggl]\,,
\nonumber\\
\frac{\partial f(z)}{\partial\overline z}+[1+f'(z)]
\xi(z)\frac{\partial\xi(z)}{\partial\overline z}=0
\label{ttt}
\end{eqnarray}
where $f'(z)=\partial_zf(z)$, $\xi'(z)=\partial_z\xi(z)$. Due to
the last relation, the resulted zweibein is
conformal flat that
follows from (\ref{tra}) along with the explicit
superbein \cite{dan90} or \cite{howe}. In this case
every $p$-supertensor receives the usual
factor $[D(t)\tilde\vartheta]^{-p}$. Ghost zero modes are
transformed, in addition, by the $\partial q_P/\partial\tilde
q_R$ matrix acting from the left.
Furthermore, $<FB>\to<FB>_\phi$ plus
an additional term \cite{dannph} due to (\ref{jump}).
In turn, it originates the
addition to $T$, which is, however, canceled (the deriving is
omitted here) by the addition to $\partial_{q_N}
\widehat{\chi}_N^{(\phi+)}(t)$. The last addition emerges since
the transition functions ({\ref{ttt}) depend on moduli. As the
result, (\ref{equ}) arises
where (\ref{bzvac}) and
(\ref{amplit}) are related as
\begin{equation}
\widehat{Z}_{L,L'}(\{q_P,\overline q_P\};\{\phi\})=
{\cal J}_L\overline{{\cal J}_{L'}}
\widetilde{Z}_{L,L'} (\{\tilde q_P,\overline{\tilde
q_P}\})\,,\quad <V>_\phi=<V>
\label{locf}
\end{equation}
where ${\cal J}_L$  is the jacobian of
the $\tilde q_R\to q_R$
transformation for the right movers, and ${\cal J}_{L'}$ is
the jacobian for the transformation of the left ones.
For simplicity we imply that the above transformations are
superholomorphic ones. Generally, for $L\neq L'$ the transformations
are distinguished from each other since the discussed transformation
depends on the spin structure. The local vacuum
expectation of the interaction vertex product in $<V>$ and in
$<V>_\phi$  differ from each other by the multiplier to be the
product of the jacobians, every jacobian being
calculated for the $(\tilde t_j,\overline{\tilde
t_j})\to (t_j,\overline t_j)$ transformation of the given interaction
vertex coordinate $t_j$.
Evidently, relations (\ref{locf}) are
just the correct change of the integrand for the module integral
under the change of the moduli. They, however, do not acheived
for the amplitude \cite{ver,as,hok} due to the discussed mistake in
\cite{ver,as,hok}. Being incorrect, the amplitude \cite{ver,as} and
\cite{hok} also do not related to each other by the jacobian of the
corresponding transformation of the moduli.

Both $\tilde t(t)$ and $\tilde q_M(\{q_N\})$
are calculated by the method developed in \cite{dannph} for
the modular transformations. In this case one starts with
obvious relations
\begin{equation}
f(z)=-\oint_z\widetilde G_b(z,z')f(z')\frac{dz'}{2i}\,,
\quad
\xi(z)=-\oint_z\widetilde G_f(z,z')\xi(z')\frac{dz'}{2i}
\label{tt0t}
\end{equation}
where the integration contour surrounds $z$. The above contour
is going to the infinity, the integrals of
discontinuities of $(f,\xi)$ emerge, as well as the terms due to
the poles at $z=z_{(i)}$. The above discontinuities
are calculated \cite{dannph} by the set of equations
$\tilde\Gamma_s(\tilde t(t))=\tilde t(\Gamma_s(t))$.
In doing so for the calculation of $(f(z),\xi(z))$ the set of the
integration equations emerge. Simultaneously $\tilde
q_M(\{q_N\})$ are calculated from the condition that
extra-discontinuities due to those (\ref{chbgr}) and
(\ref{chfgr}) of the Green functions to be canceled. The kernels
of the equations are proportional to Grassmann moduli. Although
the equations can be solved by the iteration procedure for any
number $n$ of the loops, the completeness of the expressions
rapidly increases when $n$ grows. This is the reason why the
local amplitude in the hidden supersymmetry description is
tremendous for $n>2$ although in the supercovariant scheme a
rather compact expression can be given for an arbitrary $n$.

The genus-2 case being considered,
we fix $(\tilde u_1,\tilde
v_1,\mu_1,\nu_1,\tilde u_2)$ as $\tilde u_1=u_1$, $\mu_1=0$,
$\tilde v_1=v_1$, $\nu_1=0$ and $\tilde u_2=u_2$.  Furthermore,
we define $\tilde q_m=q_m-\delta q_m$. The above mentioned
equations $\tilde\Gamma_s(\tilde t(t))=\tilde t(\Gamma_s(t))$
are reduced to the following ones
\begin{eqnarray}
\delta g_1(z)+[\partial_zg_1(z)]f(z)=f(g_1(z))\,,\quad
\sqrt{\partial_zg_2(z)}\xi(z)+
\sqrt{\partial_zg_2(z)}\epsilon_2(z)=\xi(g_2(z))
\nonumber\\
\delta
g_2(z)+[\partial_zg_2(z)][f(z)+\xi(z)\epsilon_2(z)
-\sqrt{\partial_zg_2(z)}
\frac{\mu_2\nu_2(z-g_2(z))}{u_2-v_2}=f(g_2(z))\,,
\label{tt4t}
\end{eqnarray}
where $\delta g_s(z)=\tilde g_s(z)-g_s(z)$
and $(\tilde g_s(z)|\epsilon_s(z))$ are the transition functions for
super Schottky group transformation \cite{dannph}. As it has been
noted, one transforms (\ref{tt0t}) to the integrals over
non-contractable cycles.  Using (\ref{tt4t}), one obtains that the
desired $f(z)$ and $\xi(z)$ are represented by the expressions
\begin{eqnarray}
f(z)= \frac{\lambda_1\lambda_2}{4}\biggl[\widetilde
G_b(z,z_1) \widetilde G_f (z_{(1)},z_{(2)}) -\widetilde
G_b(z,z_2)\widetilde G_f(z_{(2)}, z_{(1)})- \nonumber\\
-\oint_2\widetilde G_b(z,z')[P_1(z)
\widetilde G_f(z,z_{(2)})-
P_2(z)\widetilde{G}_f(z,z_{(1)})]\frac{dz'}{2i}\,,
\nonumber\\
\xi(z)=\lambda_1\widetilde G_f(z,z_1)+\lambda_2\widetilde
G_f(z,z_2)\,,
\label{tt1t}\\
P_i(z)=2\frac{z-v_2}{u_2-v_2}\widetilde{\chi}_{\mu_2}
(z_{(i)})-
2\frac{z-u_2}{u_2-v_2}\widetilde{\chi}_{\nu_2}(z_{(i)})
\label{tt3t}
\end{eqnarray}
provided that $\tilde q_m=q_m-\delta q_m$ are given by
\begin{eqnarray}
\delta q_m=
\frac{\lambda_1\lambda_2}{4}\biggl[\widetilde{G}_f
(z_{(1)},z_{(2)})
\widetilde{\chi}_m(z_{(1)})-\widetilde{G}_f(z_{(2)},
z_{(1)})
\widetilde{\chi}_m(z_{(2)})-
\nonumber\\
-\oint_2 dz
\widetilde{\chi}_m(z)[P_1(z)
\widetilde{G}_f(z,z_{(2)})
-P_2(z)\widetilde{G}_f(z,z_{(1)})]\biggl]\,,
\nonumber\\
\mu_2=\lambda_1\widetilde{\chi}_{\mu_2}(z_{(1)})+
\lambda_2\widetilde{\chi}_{\mu_2}(z_{(2)})\,,\,\,
\nu_2=\lambda_1\widetilde{\chi}_{\nu_2}(z_{(1)})+
\lambda_2\widetilde{\chi}_{\nu_2}(z_{(2)})
\label{tttr}
\end{eqnarray}
where $\widetilde G_b(z,z')$, $\widetilde G_f(z,z')$,
$\widetilde\chi_m(z)$ and $\widetilde\chi_\alpha(z)$ with
$\alpha=(\mu_2,\nu_2)$ are the same as in (\ref{zergr}) and
(\ref{corgr}). The integral is performed along the contour
surrounding to the positive direction both Schottky cycles
of the handle 2 and, for the Ramond
type handle, the cut between $u_2$ and $v_2$, which exists
\cite{danphr,dannph} in this case.
Furthermore,
$\widetilde{G}_f(z,z_{(2)})$
has no discontinuities around $(A_1,B_1)$-ones as far as
$(\mu_1,\nu_1)$ being fixed, are not moduli.  So
$\delta q_m$ has no discontinuities on $(z_1,z_2)$-planes.
In addition, $\widetilde{G}_f(z_{(s)},z_{(i)}) -P_i(z)$  has no
discontinuities on $z_{(s)}$-plane around $(A_2,B_2)$-cycles.
Indeed, the discontinuity
around $(A_2,B_2)$-cycles on $z_{(i)}$-plane of the integral
being due to pole of $\widetilde G_f(z,z_{(i)})$ at
$z=z_{(i)}$ coincides with the discontinuity of
$-P_s(z_{(i)})$ (the integrals of this kind
are considered in \cite{dannph}).
If the Schottky
moduli are used, then $\delta q_m$ are to be $(\delta
v_2,\delta k_1,\delta k_2)$.
If, as the moduli, one uses the
$\omega_{rs}^{(0)}/(2\pi i)$ period matrix on the
Riemann surface, then $q_m\to q_{ij}$ and $\widetilde{\chi}_m(z)
\to\widetilde{\chi}_{ij}(z)=J_i(z)J_j(z)$. In this case
$i\leq j$ and
where $J_r(z)$ is the scalar
function having periods $\omega_{rs}^{(0)}$.  As above, we
identify $q_m'$ in (\ref{ampl}) with $q_m$.

\section{Calculation of the local amplitude}

The vacuum function in (\ref{bzvac}) can be calculated from
the first of (\ref{equ}). The second of (\ref{equ}) is
satisfied, if the known interaction vertex \cite{fried} is used.
The desired vacuum function is represented as
\begin{equation}
\widehat{Z}_{L,L'}(\{q_M,\overline
q_M\};\{\phi\})= \frac{\Xi_L(\{q_m\};\{z_{(i)}\})
\overline{\Xi_{L'}(\{q_m\};\{z_{(i)}'\})}}{
{\det}^5[2Re\,\omega^{(0)}(\{q_m\})]}
\breve Z_{L,L'}(\{q_M,\overline q_M\};
\{z_{(i)},\overline z_{(i)}'\})
\label{zetzer}
\end{equation}
where $\omega^{(0)}(\{q_m\})/2\pi i$ is the ordinary period
matrix, and $\breve
Z_{L,L'}(\{q_M,\overline q_M\}; \{z_{(i)},\overline
z_{(i)}'\})=1$ when $\{\lambda_i=0\}$.  Both $\breve
Z_{L,L'}(\{q_M,\overline q_M\};\{z_{(i)},\overline z_{(i)}'\})$
and $<V>_\phi$ are calculated expanding in powers of $\phi_m$
the exponential $\exp[S_m^{(\phi)}(X)+
S_{gh}^{(\phi)}(B,F;\overline B,\overline F)]$.  One can also
calculate $<V>_\phi$ using the scalar superfield vacuum
correlator in the gravitino field, the result being expanded in
powers of  $\phi_m$.  By above, the ghost supercurrent is
calculated in terms of $(B,F)$ rather than through $(B,C)$. In
doing so eq.(\ref{corfb}) is used. An uncertainty of
$G_{3/2}(z_{(j)},z_{(s)};\{z_{(i)}\})$ at the
$(z_{(j)},z_{(s)})$ is removed  calculating
the correlator for the "spread" gravitino field.
The same result can be obtained by the
transformation (\ref{tttr}) of the amplitude (\ref{amplit})
having no the uncertainty discussed.
To choose the given $\partial
\ln\widehat{Z}_{L,L'}(\{\phi\})/\partial q_N$ in (\ref{equ}),
one multiplies both parts of the equation
by a relevant polynomial in $t$, the result
being integrated along non-contractable cycles.  This
calculation (omitted here) is very similar to that in
\cite{danphr,dannph}.
Really (\ref{equ}) is used only
to find $\Xi_L(\{q_m\};\{z_{(i)}\})$ in (\ref{zetzer}).
For doing so both part of (\ref{equ}) at zero Grassmann moduli
are multiplied by $P_m^s(z)$ in (\ref{corgr}) and the relation
(\ref{nrmr}) of Appendix A is used. Hence
$\Xi_L(\{q_m\};\{z_{(i)}\})$ satisfies to
the set of the equations, which are
differential with respect to $q_m$ where the derivatives versus
moduli is calculated with the conformal flat metric.
In this case
\begin{equation}
\Xi_L(\{q_m\};\{z_{(i)}\})=
\frac{Z_L'(\{q_m\})}
{\det[\widetilde\chi_\alpha(z_{(i)})]}
\,,
\quad
\Xi_L(\{q_m\};\{z_{(i)}\})=
\Xi_L'(\{q_m\};\{p_n\};
\{z_{(i)}\})/\det[\widetilde\chi_m(p_n)]
\label{zerep}
\end{equation}
where $Z_L'(\{q_m\})$ is given in \cite{danphr,dannph} through
the Schottky parameters (see eqs. (B5)-(B7) in \cite{dannph}
where $Z_L'(\{q_m\})$ is denoted as
$Z_{0(m)}Z_{0(gh)}H(\{q_{N_s}\})$ and $H(\{q_{N_s}\})$ is taken
at zero Grassmann parameters). It is independent of locations of
Beltrami differentials. Zero modes $\widetilde\chi_m(z)$ and
$\widetilde\chi_\alpha(z)$ were discussed above in Sec.2, see
eq.(\ref{corgr}) and the next equations. Furthermore,
it can be shown (the proof is omitted)
that $\Xi_L'(\{q_m\};\{p_n\}; \{z_{(i)}\})$ satisfies to
equations
\cite{ver} for chiral determinants.  Hence, up to a constant factor,
$\Xi_L'(\{q_m\};\{p_n\}; \{z_{(i)}\})$ is none other than the
product of the corresponding chiral determinants in \cite{ver}
with location points $\{p_n\}$ and $\{z_{(i)}\}$
given in \cite{ver} through the theta-like functions
(see eqs.
(7.5)-(7.7) in Nucl.Phys \cite{ver}).
Hence the local
amplitude (\ref{bzvac}) can be given as in terms of the
theta-like functions like \cite{ver}, so, like
\cite{danphr,dannph}, directly through Schottky parameters.  The
amplitude differs from \cite{ver} in the ghost contribution, see
Appendix C as an example. In
particular the correlator $\widetilde G_b(z,z')$ replaces
$G_2(z',z;\{p\})$ in \cite{ver} (see for definition Sec.
2 and Appendix A). In addition, unlike \cite{ver}, the module integral
contains the boundary terms discussed in the next section.

\section{Integration region and boundary terms}

The superstring amplitude  (\ref{amplit}) in the supercovariant scheme
contains no the gravitino field.
On the other side, from the previous Section, the integrand
(\ref{bzvac}) of the module integral (\ref{amplit})
depends on the gravitino field locations
$(z_{(i)},\overline z'_{(i)})$. In reality the
$(z_{(i)},\overline z'_{(i)})$-dependent terms
are total derivatives in the module space, which being
integrated by parts, are canceled by relevant boundary terms
(provided of a certain prescription to be for the integration of
singularities, see Sec. 9).

The boundary terms arise because (\ref{amplit}) is co-variant
\cite{dannph} under modular group transformations on the non-split
supermanifold \cite{bshw}. Generally, the transformation in question
presents a globally defined, holomorphic superconformal non-split
change of $\tilde t$ accompanied by a holomorphic non-split change of
$\{\tilde q_R\}$ and by the change of the spin structure, as well. The
resulted modular parameters and supercoordinates depend, however, on
the spin structure by terms proportional to Grassmann moduli. So they
are distinguished for different spin structures, if the former ones are
taken to be the same for all the spin structures discussed. As the
result, the sum over spin structures is non-covariant under modular
group on the supermanifold, if the super Schottky group moduli are
chosen to be the same for all the superspin structures. To restore the
former integral, a re-definition of the integration variables must be
performed separately for every given spin structure. This is a subtle
procedure since the integral of a single spin structure is divergent
(see \cite{danpr1} for details). In any case
the integration with respect to $\{\tilde q_M\}$
is performed over the fundamental region of the
supermodular group above. The boundary of the region is formed
by moduli related to each other by supermodular transformations,
mixing boson and fermion moduli. Like usual
modular transformation \cite{siegal}, the supermodular
transformation determines a new basis of non-contractable
cycles. In this case $\omega(\{\tilde q_M\};L)/(2\pi i)$ in
(\ref{amplit}) is changed in the same way as the ordinary period
matrix is changed \cite{siegal} under the corresponding modular
transformation of the Riemann surface. So the boundary of the
fundamental region is determined by a set of conditions
$\widehat{\cal G}_i(\omega(\{\tilde
q_M\};L),\overline{\omega(\{\tilde q_M\};L')})=0$ obtained by a
relevant superconformal extension \cite{danpr1} of the
fundamental region boundary \cite{siegal}.  Hence we present the
amplitude by the integral of the expression, which is the local
amplitude (\ref{amplit}) multiplied by the ${\cal
O}(\{\widehat{\cal G}_i\})$ cut-off factor,
\begin{eqnarray}
{\cal O}(\{\widehat{\cal G}_i\})=\prod_i\varrho(\hat{\cal
G}_i)\,,\quad\widehat{\cal G}_i\equiv \hat{\cal
G}_i(\omega(\{\tilde q_M\};L),\overline{\omega(\{\tilde
q_M\};L')})\,,
\nonumber\\
\varrho(x)=1\,\, {\rm at}\,\, x>0\,\,
{\rm and}\,\, \varrho(x)=0\,\, {\rm at}\,\, x<0\,.
\label{cutof}
\end{eqnarray}
The $\varrho(x)$ step function is understood to
be expanded over the Grassmann parameters containing in $x$ that
gives rise to desired boundary terms in the integral.  Up to the
boundary terms, the integration is performed over the
fundamental region \cite{siegal} of the ordinary modular group.
"Soul" shifts of the integration variables change the
boundary terms, too. As the result, the integral is
independent of the choice of the integration variables.
By supermodular transformations, the given integral
can be reduced to the integral over the fixed fundamental region.
So, just as it is required, the integral is independent of
the fundamental region, which is performed over
(provided that
the integration over singularities is performed
correctly, see Sec. 8).

Going to  $q_M$,
one obtains the integral (\ref{ampl}) over $q_m'=q_m$.
By above, in this case
$\omega(\{\tilde q_M\};L)/2\pi i$ in (\ref{cutof})
is calculated as the
$\hat\omega_{rs}^{(+)}(\{q_M\},\{z_{(i)}\};L)/2\pi i$
periods of the superscalar functions
for the $D_+^{(\phi)}$ operator, see Appendix
B. As an example, the genus-2 period matrix  is found to be
\begin{equation}
\hat\omega_{rs}^{(+)}(\{q_M\};\{z_{(i)}\};L)=
\omega_{rs}^{(0)}(\{q_m\})-\frac{\lambda_1\lambda_2}{2}
[\partial_{z_{(1)}}J_r(z_{(1)})]R_f(z_{(1)},z_{(2)};L)
[\partial_{z_{(2)}}J_s(z_{(2)})]
\label{2pmat}
\end{equation}
where, as above,
$\omega_{rs}^{(0)}(\{q_m\})/2\pi i$ is the
period matrix on the Riemann surface, $J_r(z)$
is the scalar
function having periods $\omega_{rs}^{(0)}(\{q_m\})$, and
$R_f(z,z';L)$ is the (1/2,1/2) Green function,
$R_f(z,z';L)\to(z-z')^{-1}$ at $z\to z'$.

Evidently, the period matrix
$\hat\omega_{rs}^{(+)}(\{q_M\},\{z_{(i)}\};L)/2\pi i$
is untouched
under the $\tilde{\cal G}_n$ group of isomorphic replacements of
the set of forming Schottky group transformations
$\{\Gamma_s\}$ by the $\{G_s\Gamma_sG_s^{-1}\}$ set. Here
$\{G_s\}$ is a relevant set of the transformations of the given
Schottky group (not every $\{G_s\}$ set originates the
isomorphism). It can be shown \cite{danpr1}) that
the space of the period matrices is covered by that region ${\cal O}'$
of the Schottky parameters where no group limiting points are
inside the common interior of any pair of Schottky circles
assigned to the forming group transformations. So, the Schottky
variables being used, the integration region ${\cal O}_m$ is, in
addition, restricted by ${\cal O}'$. Instead of
${\cal O}'$ one can use the $\widetilde{\cal O}'$ restriction, which is
an extension \cite{danpr1} of ${\cal O}'$ to the super Schottky group
description in the supercovariant gauge.
In this case $\widetilde{\cal O}'$ differs from ${\cal O}'$ solely by
boundary terms, which are canceled each by other.
Then the integration region ${\cal O}_m$
over the module space (including the boundary terms) can be given
on equal foots
though the step function product as
\begin{equation}
{\cal O}_m=
{\cal O}(\{\hat{\cal G}_i\}){\cal O}'\quad{\rm or}\quad
{\cal O}_m=\widetilde{\cal O}_m=
{\cal O}(\{\hat{\cal G}_i\})\widetilde{\cal O}'
\label{dom}
\end{equation}
where $\widetilde{\cal O}_m$ restricts the integration region in the
supercovariant description \cite{danpr1}. As it is usually, one can
replace any part of the fundamental region (\ref{dom}) by a congruent
part and still have a fundamental region. The integral is required be
independent of the fundamental region, which it is calculated over.

Only for the 2-loop and 3-loop amplitudes the boundary terms can
be removed by
the relevant $q_r\to q_r'$
replacement of the moduli.
The above replacement annihilates the dependence on
Grassmann moduli of $\hat\omega^{(+)}(\{q_M\},\{z_{(i)};L)/2\pi
i$, which is period matrix on the supermanifold. Hence
$\hat\omega^{(+)}(\{q_M\},\{z_{(i)};L)=\omega^{(0)}(\{q_m'\})$,
where, as above, $\omega^{(0)}(\{q_m'\})/2\pi i$ is the
ordinary period matrix.
Simultaneously, the integrand
becoming covariant under modular transformations on the
Riemann surface, ceases to depend on the gravitino field
locations.  We consider the 2-loop case as an example.
From (\ref{2pmat}),
\begin{equation}
q_r= q_r'+\lambda_1\lambda_2\delta
q_r'\,,\qquad \sum_r\delta q_r'
\frac{\partial\omega_{mn}^{(0)}(\{q_s'\})}{\partial q_r'}
=\frac{1}{2}[\partial_{z_{(1)}} J_m(z_{(1)})]
R_f(z_{(1)},z_{(2)};L)
[\partial_{z_{(2)}}J_n(z_{(2)})]
\label{r2mod}
\end{equation}
where $q_m'$ are treated having no soul parts.  In particular,
$\{q_m\}$ can be identified with the elements of
the $\omega^{(0)}$ matrix. Then, from (\ref{2pmat}) and
(\ref{r2mod}), it follows that $q_m'$ are identified with the
elements of the period matrix on the supermanifold.
The transition group transformations (\ref{trgrtr}) are, however,
non-split in $q_M'$ moduli that must be taken into account in the
calculation of the interaction amplitude, see Sec. 7 for more details.

The kindred consideration can be given for the genus-3 surface where
the period matrix is again isomorphic to the boson moduli. For the
genus-$n>3$ one can not remove Grassmann moduli from
$\hat\omega^{(+)}(\{q_M\},\{z_{(i)};L)$ ever so the module slice is
used. Simultaneously, the above terms depend on the gravitino field
locations.
One can check the above statements, for instance, for $n=4$. For doing
so one calculates $\hat\omega_{rs}^{(+)}(\{q_M\},\{z_{(i)}\};L)/2\pi
i$ by the substitution $\tilde q_P=\tilde (\{q_R\})$ into
$\omega(\{\tilde q_M\};L)/2\pi i$. The $\omega(\{\tilde q_M\};L)/2\pi
i$ period matrix is given in \cite{danphr,danpr1}. In turn, $\tilde
q_P(\{q_R\})$ is calculated for the transformation (\ref{ttt}) as it
has been discussed in Section 3. Hence for $n>3$ boundary terms
are necessary present. In addition, they are dependent on the gravitino
field locations and on the spin structure, as well.  The integrand
depends on the gravitino field locations by terms, which are total
derivative of a local function of the boson moduli once the integration
versus Grassmann moduli has been performed.  Naively, the discussed
terms are canceled by corresponding pieces of the boundary terms
(really the integration is ill defined due to singularities, see Sect.
8).  Nevertheless, for $n>3$ the boundary terms are necessary remained
although they are independent of the gravitino field locations.  It is
the evidence that the rest integrand receives an additional term under
the modular transformation. Just the bad modular property of the
integrand does not allow to obtain the vacuum amplitude vanishing
locally in the module space for the number of loops to be $n>3$ (and
for the 2- and 3-loop case with the usual choice \cite{ver} of the
module slice).  Modular transformation discussed can be obtained from
(\ref{locf}) using  the co-variance of $\widetilde{Z}_{L,L'} (\{\tilde
q_P,\overline{\tilde q_P}\})$ under the modular transformations on the
$\tilde t$ supermanifold (see Appendix C the two-loop amplitude, as an
example).

\section{Two-loop amplitude}

Now we consider the calculation of the 2-loop amplitude
for the moduli slice where the boundary integral is absent. In
particular, we give the correct expression
for the vacuum amplitude instead of the mistakable one in \cite{hok}.
In this section we denote the integrand $\widehat
F_{L,L'}(\{q_M,\overline q_M\};\{\phi\})$ in (\ref{ampl}) as
$\widehat B_{L,L'}^{(r)}$. Only the
$\sim\lambda_1\lambda_2\overline\lambda_1\overline\lambda_2$
term
$\lambda_1\lambda_2\overline\lambda_1\overline\lambda_2
\widehat B_{L,L'}^{(r11)}$ of
$\widehat B_{L,L'}^{(r)}$ contributes to
the integral.
To see that $\widehat B_{L,L'}^{(r11)}$
is independent of $\{z_{(i)},\bar z_{(i)}'\}$,
it is convenient to obtain  it
through the supercovariant gauge amplitude.
For doing so one
turns in (\ref{tbhb}) of Appendix C
from the moduli $q_m$ to
$q_m'$ related with $q_m$ by (\ref{r2mod}).
So
\begin{eqnarray}
\widehat B_{L,L'}^{(r11)}=
\Biggl[\widetilde{{\cal B}}_{L,L'}^{(11)}+
\frac{\partial}{\partial q_m'}\frac{\partial}{\partial
\overline q_n'}\biggl[{\cal K}_m(L)\overline{\cal
K}_n(L')\widetilde{{\cal B}}_{L,L'}^{(00)}\biggl]
-\frac{\partial}{\partial q_m'}\biggl[{\cal K}_m(L)
\widetilde{{\cal B}}_{L,L'}^{(01)}\biggl]
-\frac{\partial}{\partial
\overline q_n'}\biggl[\overline{\cal
K}_n\widetilde{{\cal B}}_{L,L'}^{(10)}\biggl]\Biggl]\,,
\nonumber\\
{\cal K}_m=\frac{\delta q_m-\lambda_1
\lambda_2\delta q_m'}
{\det[\widetilde\chi_\alpha(z_{(i)})]}\quad
\label{bfun}
\end{eqnarray}
where definitions are the same as in (\ref{amplit}), (\ref{tttr}),
(\ref{zerep}), (\ref{r2mod}) and (\ref{tbhb}). In (\ref{bfun}) a
dependence on $(z_{(1)}, z_{(2)})$ might be only due ${\cal K}_m$.  One
can, however, see that the residue at $z_{(1)}\to z_{(2)}$ on the right
side of (\ref{r2mod}) is $[\partial_{z_{(2)}}J_m(z_{(2)})]
[\partial_{z_{(2)}}J_n(z_{(2)})]$ that is
$[\partial_{q_r'}\omega_{mn}^{(0)}(\{q_s'\})]
\widetilde{\chi}_r(z_{(2)})$. To verify it, one can take the moduli
to be $\omega_{mn}^{(0)}$. Then the ghost zero modes just are
$[\partial_{z_{(2)}}J_m(z_{(2)})]
[\partial_{z_{(2)}}J_n(z_{(2)})]$, see the text next to
(\ref{tttr}). So $[\delta q_m'-\delta\hat q_m]$ is finite at
$z_{(1)}=z_{(2)}$. Moreover, $[\delta q_m-\delta\hat q_m]$
is anti-symmetrical in its arguments and is $3/2$-tensor in each
one of them, it depends on $(z_{(1)},z_{(2)})$ only by the
$\det[\widetilde\chi_\alpha(z_{(i)})]$ factor that is the
denominator of ${\cal K}_m$.  Hence ${\cal K}_m$ does not depend
on $(z_{(1)},z_{(2)})$, and, therefore, $\widehat
B_{L,L'}^{(r11)}$ is independent of $(z_{(1)},z_{(2)})$, too.
As far as the supermanifold period matrix in the $q_p'$ moduli does not
depends on the Grassmann ones, the vacuum function is the product of
the factorized expression times the usual non-holomorphic factor. So
\begin{equation}
\widehat B_{L,L'}^{(r)}=
\frac{Z_L(\{q_M'\};\{z_{(i)}\})
\overline{Z_{L'}(\{q_M'\};\{z_{(i)}'\})}}{
{\det}^5[2Re\,\omega^{(0)}(\{q_m'\}]}
<V>_r
\label{tiloc}
\end{equation}
where $<V>_r$ is
the vacuum expectation $<V>_\phi$  of $V$ in (\ref{bzvac}) where
the $q_r$ moduli are expressed
by (\ref{r2mod}}). The vacuum function
differs from that in (\ref{bzvac}) only in terms
proportional to Grassmann moduli. So
\begin{equation}
Z_L(\{q_M'\};\{z_{(i)}\})=\Xi_L(\{q_m'\};\{z_{(i)}\})
[1-\lambda_1\lambda_2Z_L^{(r)}(\{q_m'\};\{z_{(i)}\})]\,,
\label{vc}
\end{equation}
where $\Xi_L(;\{z_i\})$ is the same as
in (\ref{zetzer}). In the considered case the fermion
moduli do not agitated to the boson ones under the supermodular
transformations. Hence
(\ref{tiloc}) is covariant under the modular group on the Riemann
surface. In particular,  the zero point function
\begin{equation}
Z=\sum_L\Xi_L(\{q_m\};\{z_i\})Z_L^{(r)}(\{q_m\};;\{z_{(i)}\})/
\det[\partial\omega_{(j)}/\partial q_m]
\label{zpf}
\end{equation}
is
invariant under the modular transformations. Here
$\omega_{(1)}=\omega_{11}(\{q_r\})$,
$\omega_{(2)}=\omega_{22}(\{q_r\})$ and
$\omega_{(3)}=\omega_{12}(\{q_r\})$.
In addition, by above, (\ref{zpf}) is independent of
$\{z_{(i)}\}$. Thus one can prove that $Z=0$.
Indeed, using relation
(\ref{bfun}) for the vacuum function and the explicit
expressions \cite{danphr,danpr1,dannph} of the vacuum function
one can derive
that $Z$ has no a singularity in the $(k_1,k_2)$ Schottky
multipliers when either $k_1\to0$, or $k_2\to0$.  In this case due to
the modular symmetry, $Z$ has no singularity in each of $k_s$ on the
complex $k_s$-plan. Thus $Z$ independent of $k_s$.  Being modular
invariant, $Z$ is independent of $v_2$, too. Otherwise it ought to
receive a dependence on the Schottky multipliers under the relevant
modular transformation since $v_2$, generally, depends on the resulted
Schottky multipliers. On the other hand, at $v_2\to u_2$ it is the
product of the torus vacuum function by the torus vacuum one, each
being nullified.  So $Z\equiv0$ identically.
At the same time, as wee shall see, $Z_L^{(r1)}(\{q_m'\})$ is
different from the corresponding expression in \cite{hok}.
It is convenient to calculate it from
(\ref{bzvac}) by using the transformation (\ref{r2mod}).
Using eq.(\ref{vacfn}) from Appendix C, one obtains
that
\begin{eqnarray}
Z_L^{(r1)}(\{q_m'\})= \frac{\partial\delta
q_m'}{\partial q_m'}+\delta q_m' \frac{\partial}{\partial
q_m'}\ln\Xi_L(\{q_m'\};\{z_i\})+
5R_f(z_1,z_2)\partial_{z_1}\partial_{z_2}
R_b(z_1,z_2)+
\nonumber\\
+\widetilde W_L(z_1,z_2)-\widetilde
W_L(z_2,z_1)\,,
\nonumber\\
\widetilde
W_L(\hat z_1,\hat z_2)
=\widetilde{G}_b(\hat
z_2,\hat z_1)\biggl[\partial_{z}G_{3/2}(z,z_1)
\biggl]_{z=\hat z_2}
-\frac{1}{2}\biggl[\partial_z\chi_{\hat
z_1}(z)\biggl]_{z=\hat z_2}
\partial_{\hat z_1}\widetilde{G}_b(\hat z_2,\hat z_1)
\label{parti}
\end{eqnarray}
where $\Xi_L(\{q_m'\};\{z_i\})$ is defined in
(\ref{zetzer})), $R_b(z,z')$ is the scalar Green function having
discontinuities, and other definitions are in
(\ref{zergr})-(\ref{fgrz}), (\ref{2pmat}) and (\ref{r2mod}).
The terms in (\ref{vacfn}) with the derivatives versus locations
are canceled with the corresponding terms from the jacobian of
the considered transformation (it can be verify by checking that
the considered expression has no singularity, the proving being
omitted here). So in (\ref{parti}) all
derivatives versus moduli are calculated with fixed
$(z_1,z_2)$. As above, in doing so
the metric is kept to be conformal flat (generally, it is
not the same as the discussed in \cite{ver} derivatives
due to the change of the metric.
In this case
$\partial_{q_m'}\Xi_L(\{q_m'\};\{z_i\})$ is expressed in terms
of the right side of the first of (\ref{equ}), and
$\partial_{q_m'}\delta q_m'$ is calculated from its
discontinuities  in $(z_{(1)},z_{(2)})$ by the method
\cite{danphr}.
For the brevity the explicit
expressions of the derivatives are omitted here.
If the moduli are chosen to be
$\omega_{mn}^{(0)}$, then the ghost zero modes
$\widetilde{\chi}_m(p_n)$ in $\Xi_L(\{q_m'\};\{z_i\})$, see
(\ref{zerep}), are $[\partial_{z_{(2)}}J_m(z_{(2)})]
[\partial_{z_{(2)}}J_n(z_{(2)})]$, see the text next to
(\ref{tttr}).  On can
easy check that the (\ref{parti}) has no the pole at
$z_{(1)}=z_{(2)}$, it is anti-symmetrical in its arguments and
is $3/2$-tensor in each one of them. To see that
(\ref{parti}) depends on $(z_{(1)},z_{(2)})$  by the
$\det[\widetilde\chi_\alpha(z_{(i)})]$ factor,
it is yet necessary to verify for no to be the pole at the
point where $\det[\widetilde\chi_\alpha(z_{(i)})]=0$ and
$z_{(1)}\neq z_{(2)}$. It is a difficult task hampering
the check discussed. As in \cite{hok}, the expression
(\ref{parti}) is simplified, if the locations satisfy
to the condition $R_f(z_{(1)},z_{(2)};L)=0$, and, therefore,
$\delta q_m'=0$. Nevertheless, it is
different from \cite{hok}. In particular, (\ref{parti}) is
expressed through $\widetilde G_b(z_{(1)},z_{(2)})$ instead of
$G_2(z_{(2)},z_{(1)}))$.

The interaction local amplitude (\ref{tiloc}) contains, in
addition, the $<V>_r$ factor. The desired $\widehat
B_{L,L'}^{(r11)}$ function (\ref{bfun}) is the integral versus
the interaction vertex coordinates on the supermanifold,
the integrand being
dependent on the gravitino field locations due the difference
(\ref{ttt}) between $(\tilde z|\tilde\vartheta)$ and
$(z|\vartheta)$. The
above dependence
disappears once the integration versus the vertex
coordinates has been performed.
Using \cite{danphr} one can check that the  local vacuum amplitude
in the leading approximation is factorized at
$v_2\to u_2$ by the product of the genus-1 functions. So it is
plausible that the 1-, 2- and 3-point $\widehat B_{L,L'}^{(r11)}$
function is nullified and, therefore, the interaction amplitudes are
finite. In fact, however, before to conclude for this, the leading
corrections need to be examined, too.

Calculating $\widehat B_{L,L'}^{(r11)}$
directly from the field integral in (\ref{ampl}) one
needs to remember that all the correlators and zero
modes have usual form, if they are given through the
$q_r$ moduli, but the $q_r'$ ones.
Any correlator
$K(z,z';\{q_r'\})$ is
$K(z,z';\{q_r'\})=K_0(z,z';\{q_r'\})+
\lambda_1\lambda_2\delta q_r'\partial_{q_r'}
K_0(z,z';\{q_r'\})$ where $K_0(z,z';\{q_r'\})$ is its usual
expression. The above is true for zero modes, as well.
So, the calculation of the
local interaction amplitude is tremendous.
Only the vacuum function can be calculated in a rather simple
way.

\section{Berezin integral}

The results of the previous Section show that the local
amplitude can be finite
or divergent depending on the module slice.  For instance,
the 2-loop vacuum function for Verlinde-like moduli
is non-integrable (see eq.(\ref{vacfn}) from Appendix C) while
it is zero, if the module slice (\ref{ttt}) is used. It
is a particular manifestation of the fact that the same
integral with respect to both the local variables and Grassmann
ones can be found to be finite or divergent, as this is seen for
an easy integral
\begin{equation}
I_{(ex)}=\int\frac{dxdyd\alpha
d\beta d\bar\alpha d\bar\beta}
{|z-\alpha\beta|^p}\varrho(1-|z|^2)
\label{examp}
\end{equation}
where $z=x+iy$, $(\alpha,\beta)$ are Grassmann
variables,
$p$ characterizes the
strength of the singularity and $\varrho$ is
the step function, see (\ref{cutof}).
Singular integrals of this kind really
arise in (\ref{ampl}). Indeed,
Schottky variables being employed,
the integration measure
contains \cite{danphr,dannph} the singularity as
in (\ref{examp}) at $p=2$. For the sake of simplicity we bound
the integration region taking $|z|^2\leq1$. Moreover, it is
usually to re-write down (\ref{examp}) as
\begin{equation}
I_{(ex)}=\int\frac{dxdy}
{|z|^p}\varrho(1-|z|^2)\,d\alpha\,
d\beta\,d\bar\alpha\,d\bar\beta+
p^2\int\frac{dxdy} {4|z|^{p+2}}\varrho(1-|z|^2)
\alpha
\beta\bar\alpha\bar\beta\,
d\alpha\,
d\beta\,d\bar\alpha\,d\bar\beta
\,,
\label{exam}
\end{equation}
where the first integral on the right side is naively treated to
be zero since it does not explicitly contain Grassmann
variables.  The second integral is divergent at $z=0$, if
$Re\,p>0$. On the other side, in (\ref{examp}) one can introduce
$\tilde z=z-\alpha\beta$ instead of $z$. Then the Grassmann
variables will be only in the step function $\varrho(|\tilde
z+\alpha\beta|^2)$. Once the Grassmann integrations to be
performed, the integral is reduced to the integral along the
circle $|z|^2=1$. Thus for any $p$ the result is
finite as follows
\begin{equation}
I_{(ex)}=-\int\frac{d\tilde
xd\tilde yd\alpha d\beta d\bar\alpha d\bar\beta} {|\tilde
z|^p}\alpha\beta\bar\alpha\bar\beta \left[\delta(|\tilde
z|^2-1)+|\tilde z|^2\frac{d\delta(|\tilde z|^2-1)}{d|\tilde
z|^2}\right]=-\frac{\pi p}{2}\,.
\label{ex}
\end{equation}
So (\ref{examp}) depends on the integration
variables, at least for $Re\,p>0$, see \cite{danpr1} for more
details. The reason is that the integrand is expanded in a
series over the Grassmann variables even though it is singular.
It is commonly the integral of a singular expression
to define with a cut-off, which excludes a
small domain containing the singular
point. Thus (\ref{exam}) is calculated with the cut-off
$\varrho(|z|^2-o)$ while (\ref{ex}) is calculated with
$\varrho(|\tilde z|^2-o)$ both at $o\to0$.
In the last case the first
integral on the right side of (\ref{exam}) is not zero since it
contains the Grassmann variables due to the $\varrho(|\tilde
z|^2-o)$ cut-off. Moreover, it is divergent while
the sum over two integral on the right side (\ref{exam})
is finite.  This just occurs with the calculation of the 2-loop
superstring amplitude in the previous Section when one starts with
its expression given (see Appendix C) through Verlinde-like moduli.
In this case naively only $\widehat{{\cal B}}_{L,L'}^{(11)}$
in (\ref{exlam}) contributes to the integral, the integral being
divergent. Once the replacement (\ref{r2mod}) has been
performed, $\widehat{{\cal B}}_{L,L'}^{(11)}$ is added by pieces
arising in the rest terms of (\ref{exlam}), the integral of the
resulting expression (\ref{tiloc}) being finite.

In the general $n>3$ case one can remove divergences by a change of the
moduli, but, as it has been discussed in Sec. 6, the boundary terms
always present, and the local amplitude is not covariant under the
modular transformations on the Riemann surface. Thus the desired
moduli are not unique and the result may depend on the moduli employed.
A relevant choice of the moduli is that, which ensures the local
symmetries of the amplitude. In particular, the result is required to
be the same for the fundamental region, which the local amplitude is
integrated over, boundary terms being included. As far as the boundary
terms depend on the gravitino field locations $\{z_{(i)},\bar
z_{(i)}\}$, the above requirement leads, among other things, to the
condition that the whole
amplitude is independent of $\{z_{(i)},\bar z_{(i)}\}$.  For $n>3$
the above program seems, however, to be
inconvenient for the actual calculation of the amplitude.

Another way has deal directly with a divergent integral. In this
case one can try to extract the gravitino field dependence of the local
amplitude in the form of the derivatives versus the moduli and
interaction vertex coordinates of the relevant local functions
once the Grassmann integrations has been performed. The above functions
have non-integrable singularities. So an additional prescription
for the integration of the singularities is necessary in line
with the above requirement that the integral is the same for any
fundamental region chosen to be the integration one.
An actual method to extract the discussed terms
is to use the relation (\ref{locf}) between the
Verlinde-like amplitude and the amplitude given in the supercovariant
gauge.  As an example, the
2-loop Verlinde-like vacuum amplitude (\ref{vacfn}) from  Appendix C
being considered, one hardly can disply the discussed terms. They
are, however, evident from (\ref{exlam}) along with (\ref{tbhb}) and
(\ref{tbhbn}). Once the integration versus Grassmann moduli has been
performed, only (\ref{tbhb}) contributes to the integral. The
correect prescription is to remove
the total derivative terms, the integral of
$\widetilde{{\cal B}}_{L,L'}^{(11)}$ being remained. The
remaining integral is none other than the integral arising in the
supercovariant gauge once the integration versus Grassmann moduli has
been performed. Thus it is much more easy to perform the
calculation \cite{danpr1} directly in the supercovariant gauge.

\appendix
\def\thesection{Appendix \Alph{section}}
\def\theequation{\Alph{section}.\arabic{equation}}
\setcounter{equation}{0}

\section{Relations for ghost Green functions}

For instance, $\widetilde
G_b(z,z')$ and the correlator $G_2(z',z;\{p\})$ depending on
$(3n-3)$ location points $\{p_m\}\equiv\{p\}$, are related by
\begin{eqnarray}
\widetilde G_b(z,z')=G_2(z',z;\{p\}) -\int_{s}
G_2(\tilde z,z;\{p\}) \frac{d\tilde z}{2i}P_m^s(\tilde z)
\widetilde\chi_m(z')\,,
\nonumber\\
G_2(z',z;\{p\})=\widetilde{G}_b(z,z')-
\widetilde G_b(z,p_j)\chi_j(z';\{p\})\,,
\nonumber\\
\chi_j(z;\{p\})=\int_{s}\chi_2^{(j)}(\tilde z;\{p\})
\frac{d\tilde z}{2i}
P_m^s(\tilde z)
\widetilde{\chi}_m(z)\,,
\nonumber\\
\widetilde{\chi}_m(z)=\widetilde A_{mj}\chi_j(z;\{p\})\,,\quad
\widetilde A_{jm}^{-1}\widetilde A_{ml}=\delta_{ml}\,,
\quad \widetilde A_{jm}=\widetilde\chi_m(p_j)
\label{corgr}\\
G_2(p,z;\{p\})=0\,,\quad
(z'-p_j)G_2(z,z';\{p\})=
\chi_j(z;\{p\})\,\,
{\rm at}\,\,
z\to p_j\in\{p\};
\nonumber\\
\chi_j(p_l;\{p\})=\delta_{jl}\,,
\,\,p_l\in\{p\}\,.
\label{correl}
\end{eqnarray}
Here $g_s(z)\equiv g_s(z;\{q_n\})$ and
$P_m^s(z)=[\partial_{q_m}
g_s(z)]/[\partial_z
g_s(z)]$. In this case $\pi
G_2(z',z;\{p\})\to(z-z')^{-1}$ at $z'\to z$.
To derive
the first of (\ref{corgr}) one represents
$\widetilde G_b(z,z')$
by the contour integral
around $\tilde z=z$
of $-G_2(\tilde z,z;\{p\})
\widetilde G_b(\tilde z,z')$.
By moving the
contour, the integral is reduced to the integrals along
the cycles.
The discontinuity $\widetilde G_b(z,z')$ is
due to $g_s(z)$ lies
outside the fundamental region on $z$-plan, if $z$ is inside the
region above. The  relation between
zero modes provides the
cancellation of the poles
at $z=p\in\{p\}$ for $\widetilde{G}(z,z')$. It is useful to note that
\begin{equation}
\int_{s}
\frac{d\,z}{2i}\widetilde{\chi}_m(z)P^s_n(z)
=\delta_{mn}
\label{nrmr}
\end{equation}
where the integral is along the Schottky circle assigned to the handle
$s$. The boundary of the second circle is obtained by the
$z\to g_s(z;\{q_n\})$  transformation of the boundary of the first one.
In the kindred way, $\widetilde
G_f(z,z')$ and the correlator $G_{3/2}(z',z;\{p'_\alpha\})$
depending on
$(2n-2)$ location points $\{p'_\alpha\}$ are related by
\begin{eqnarray}
G_{3/2}(z',z;\{p'_\alpha\})=\widetilde G_f(z,z')`-
\widetilde G_f(z,p'_\beta)\chi_\beta(z')\,,\quad
\chi_\beta(z')=
\hat
A_{\beta\beta'}^{-1}\widetilde\chi_{\beta'}(z')\,,
\nonumber\\
\hat
A_{\beta\beta'}^{-1}\hat A_{\beta'\alpha}=
\delta_{\beta\alpha}\,,\,\,
\hat
A_{\alpha\beta}=\widetilde\chi_\alpha(p'_\beta)
\label{fgrz}
\end{eqnarray}
where $\{p_\alpha'\}$ are $(2n-2)$ location points for the
corresponding $G_{3/2}(z',z;\{p_\alpha\})$ correlator
\cite{ver}. In this case
$\pi G_{3/2}(z',z;\{p_\alpha\})\to(z-z')^{-1}$.
One can see from (\ref{corgr}) and (\ref{fgrz})
that "non-physical" poles \cite{ver} in $\{p_m\}$ and in
$\{p_\alpha'\}$ of the correlators coincides with "non-physical"
zeros of $\det\widetilde A$ in (\ref{corgr}) and, respectively,
of $\det\hat A$ in (\ref{corgr}) when the corresponding locations
all are different from each other.  Like (\ref{correl}),
\begin{eqnarray}
G_{3/2}(p,z;\{p_\alpha\})=0\,,\quad
(z'-p_{\alpha_j})G_{3/2}(z,z';\{p_\alpha\})=
\chi_{\alpha_j}(z;\{p_\alpha\})
\nonumber\\
{\rm at}\,\,
z\to
p_{\alpha_j}\in\{p_\alpha\};\quad
\chi_{\alpha_j}
(p_{\alpha_l};\{p_\alpha\})=\delta_{jl}\,,
\label{corrf}\\
\widetilde G_f(z,z')=G_{3/2}(z',z;\{p_\alpha\}) -\int_{s}
G_{3/2}(z',z;\{p_\alpha\})\frac{d\tilde
z}{2i}P_{\alpha_j}^s(\tilde z)\widetilde\chi_{\alpha_j}(z').
\label{corgrf}
\end{eqnarray}

\section{Scalar superfield correlator}

\setcounter{equation}{0}

We define
$R_\phi^{(+)}(t,t';L)/\pi$ and $R_\phi^{(-)}(\bar
t,\bar t';L')/\pi$ to be holomorphic and, respectively,
anti-holomorphic Green functions for $D_+^{(\phi)}D_-^{(\phi)}$.
In doing so
\begin{equation}
R_\phi^{(+)}(\Gamma_s(t),t';L)=
R_\phi^{(+)}(t,t';L)+J_m^{(+)}(t';\{\phi_-\};L)
\label{dhgr}
\end{equation}
where $J_m^{(+)}(t;\{\phi_-\};L)$ is the scalar function
having the periods $\omega_{sm}^{(+)}(\{q_M\};\{\phi_-\};L)$,
as follows
\begin{equation}
J_m^{(+)}(\Gamma_s(t);\{\phi_-\};L)=
J_m^{(+)}(t;\{\phi_-\};L)+2\pi i
\omega_{sm}^{(+)}(\{q_M\};\{\phi_-\};L)
\label{dscl}
\end{equation}
The Green function $R_\phi^{(+)}(t,t')$ is calculated by
\begin{equation}
R_\phi^{(+)}(t,t')=R(t,t';L)
-\int\frac{d^2 t_1}{\pi}D(t_1)R(t,t_1)
[D_+^{(\phi)}-\bar D]R_{(\phi)}^{(+)}(\tilde t,t')
\label{ehgr}
\end{equation}
where the operators are defined by (\ref{derphi}), and
\begin{equation}
R(t,t';L)=R_b(z,z')-\vartheta\vartheta'R_f(z,z';L)
\label{szgr}
\end{equation}
is $R_\phi^{(+)}(t,t')$ at $\phi=0$. Furthermore,
$R_b(z,z')\to\ln(z-z')$ and $R_f(z,z';L)\to(z-z')^{-1}$ at $z\to
z'$. Scalar superfunctions $J_m^{(\pm)}(t)$ are calculated by
(\ref{dhgr}) once $R_\phi^{(\pm)}(t,t')$ are known.  Then one
calculates the period matrix. The kindred relations take place
for the anti-holomorphic functions, as well. The desired scalar
superfield vacuum correlator $X_\phi(t,\bar t;t',\bar t')$ is
given by
\begin{eqnarray}
-4X_\phi(t,\bar t;t',\bar
t')=R_\phi^{(+)}(t,t';L) +R_\phi^{(-)}(\bar t,\bar
t';L')
\nonumber\\
-[J_m^{(+)}(t)+J_m^{(-)}(\bar
t)][\omega^{(+)}+\omega^{(-)}]_{mn}^{-1}
[J_n^{(+)}(t';)+J_n^{(-)}(\bar
t')]
\label{corx}
\end{eqnarray}
where
$J_m^{(\pm)}(t)\equiv J_m^{(\pm)}(t;\{\phi_-\};L)$, and
$\omega^{(\pm)}$ is the period matrix for the $D_{\pm}^{(\phi)}$
operator. In the
Schottky representation $R_b(z,z')$ and
$R_f(z,z';L)$ are in \cite{danphr}. In \cite{ver} both they are
given in terms of theta-functions. The scalar Green function in
\cite{ver} is different from $R_b(z,z')$, but
the difference does not contributes to the amplitude.

\section{Two-loop Verlinde-type amplitude}

\setcounter{equation}{0}

In
particular, it has been found that
\begin{eqnarray}
\frac{\breve Z_{L,L'}(\{q_M,\overline
q_M\}; \{z_i,z_i'\})\pi^{10}}{{\det}^5[2Re\,\omega(\{q_m\})]}=
\int d^{10}h_1d^{10}h_2\exp[2Re\,\omega_{il}(h_ih_l)]
\biggl[1-\frac{\lambda_1\lambda_2}{2}W_L(h)\biggl]
\nonumber\\
\times
\biggl[1-\frac{\overline\lambda_1
\overline\lambda_2}{2}\overline{W_{L'}(h)}\biggl]\,,
\label{holint}
\end{eqnarray}
\begin{eqnarray}
W_L(h)=5R_f(z_1,z_2)\partial_{z_1}\partial_{z_2}
[R_b(z_1,z_2)-(h_ih_l)J_i(z_1)J_l(z_2)]
\nonumber\\
+
\widehat W_L(z_1,z_2)
-\widehat
W_L(z_2,z_1)\,,
\nonumber\\
\widehat W_L(z_1,z_2)=\widetilde W_L(z_1,z_2)
+\widetilde\chi_m(z_1)\biggl[\partial_{z}G_{3/2}(z_1,z)
\biggl]_{z=z_2}\frac{\partial z_2}{\partial q_m}
\nonumber\\
-\frac{1}{2}\biggl[\partial_z\chi_{z_1}(z)\biggl]_{z=z_2}
\partial_{z_1}\widetilde \chi_m(z_1)
\frac{\partial z_2}{\partial q_m}\,,
\nonumber\\
\widetilde W_L(z_1,z_2)
=\widetilde{G}_b(z_2,z_1)\biggl[\partial_{z}G_{3/2}(z_1,z)
\biggl]_{z=z_2}
-\frac{1}{2}\biggl[\partial_z\chi_{z_1}(z)\biggl]_{z=z_2}
\partial_{z_1}\widetilde{G}_b(z_2,z_1)\quad
\label{vacfn}
\end{eqnarray}
where $h_s=\{h_s^{\cal M}\}$ is 10-momentum,
$(h_ih_l)$ is 10-scalar product, and other definitions are in
(\ref{zergr}) and in the accompanying it text.
The first term of $W_L(z_1,z_2)$ is due to the string
fields, and $\widehat W_L(z_1,z_2)-\widehat
W_L(z_2,z_1)$  is due to the ghost fields.
The contribution to the partition function from the first
term of $W_L(z_1,z_2)$ multiplied by
$\Xi_L(\{q_m\};\{
z_i\})\det[\widetilde{\chi}_m(p_m)]$
is the same as in \cite{ver} (this is just the first  term
inside the square brackets in eq.(38) of \cite{ver}) because, by
above,  $\Xi_L(\{q_m\};\{
z_i\})\det[\widetilde{\chi}_m(p_m)]$ is the same as the
of the chiral determinants in \cite{ver}
calculated with the fixing points to be $\{p_m\}$ and
$\{p_\alpha\}=\{z_i\}$. At the same time, the
ghost part of $W_L(z_1,z_2)$ is quite
different from \cite{ver} even though  the locations are
independent of the moduli.

To see properties of the 2-loop amplitude in question, it is useful
to represent it through the local amplitude in the
supercovariant gauge \cite{danphr,danpr1,dannph} by
the transformation (\ref{tt1t})--(\ref{tttr}).
We consider the amplitude (\ref{amplit})
multiplied by the cut-off factor (\ref{dom}). In this case
\begin{equation}
\widehat F_{L,L'}(\{q_M,\overline
q_M\};\{\phi\}){\cal O}_m=\widehat{{\cal B}}_{L,L'}^{(00)}+
\lambda_1\lambda_2
\widehat{{\cal B}}_{L,L'}^{(10)}+\overline\lambda_1\overline\lambda_2
\widehat{{\cal B}}_{L,L'}^{(01)}+
\lambda_1\lambda_2\overline\lambda_1\overline\lambda_2
\widehat{{\cal B}}_{L,L'}^{(11)}
\label{exlam}
\end{equation}
where $\widehat{{\cal B}}_{L,L'}^{(jl)}$ is independent of
$(\lambda_j,\overline\lambda_j)$. In the kindred way,
for the amplitude (\ref{amplit}) multiplied by
$\widetilde{\cal O}_m$ in (\ref{dom}), one has
\begin{equation}
\widetilde F_{L,L'}(\{q_M,\overline
q_M\};\{\phi\})\widetilde{\cal O}_m=\widetilde{{\cal
B}}_{L,L'}^{(00)}+\mu_2\nu_2 \widetilde{{\cal
B}}_{L,L'}^{(10)}+\overline\mu_2\overline\nu_2
\widetilde{{\cal B}}_{L,L'}^{(01)}+
\mu_2\nu_2\overline\mu_2\overline\nu_2
\widehat{{\cal B}}_{L,L'}^{(11)}
\label{exmunu}
\end{equation}
The desired relations are found to be
\begin{eqnarray}
\widehat{{\cal B}}_{L,L'}^{11}=
\widetilde{{\cal B}}_{L,L'}^{(11)}+
\frac{\partial}{\partial q_m}
\frac{\partial}{\partial\overline q_n}
\biggl(\delta q_m'\delta\overline q_n'
\frac{\widetilde{{\cal B}}_{L,L'}^{(00)}}{M_L\overline M_{L'}}
\biggl)
-\frac{\partial}{\partial q_m}
\biggl(\delta q_m'
\frac{\widetilde{{\cal B}}_{L,L'}^{(01)}}{M_L}\biggl)-
\frac{\partial}{\partial\overline  q_m}
\biggl(\delta\overline q_m'
\frac{\widetilde{{\cal B}}_{L,L'}^{(10)}}{\overline M_{L'}}\biggl)
\,,
\label{tbhb}\\
\widehat{{\cal B}}_{L,L'}^{10}\overline M_{L'}=
\widetilde{{\cal B}}_{L,L'}^{(10)}
-\frac{\partial}{\partial q_m}
\biggl(\delta q_m'
\frac{\widetilde{{\cal B}}_{L,L'}^{(00)}}{M_L}\biggl)\,,
\quad
\widehat{{\cal B}}_{L,L'}^{01}M_{L}=
\widetilde{{\cal B}}_{L,L'}^{(01)}
-\frac{\partial}{\partial\overline  q_m}
\biggl(\delta\overline q_m'
\frac{\widetilde{{\cal B}}_{L,L'}^{(00)}}{\overline M_{L'}}\biggl)\,,
\nonumber\\
\widehat{{\cal B}}_{L,L'}^{00}M_{L}\overline M_{L'}=
\widetilde{{\cal B}}_{L,L'}^{(00)}\,,
\label{tbhbn}\\
M_L=\widetilde{\chi}_{\mu_2}(z_1)\widetilde{\chi}_{\nu_2}(z_2)
-\widetilde{\chi}_{\mu_2}(z_2)\widetilde{\chi}_{\nu_2}(z_1)\,,
\label{matl}
\end{eqnarray}
where
$\delta q_m=\lambda_1\lambda_2\delta q_m'$, and $\delta
q_m$ is given by (\ref{tttr}).
Other definitions are the same as in (\ref{amplit}), (\ref{exlam})
and in (\ref{exmunu}).
Since in (\ref{tbhb}) the derivatives act on the cut-off as
well, the integral of the terms with derivatives naively
vanishes. Then
the right side of (\ref{tbhb}) is reduced to
$\widetilde{{\cal B}}_{L,L'}^{(11)}$ that is
independent of the gravitino field locations.
Nevertheless, it is not covariant under modular
transformations. Indeed,
(\ref{exmunu}) is covariant \cite{dannph} under the
modular transformations on the supermanifold \cite{bshw} mixing
fermion
moduli to the boson ones that originates an addition to
$\widetilde{{\cal B}}_{L,L'}^{(11)}$.


\begin{thebibliography}{99}
\bibitem{bshw}
M.A. Baranov and A.S.
Schwarz, Pis'ma ZhETF 42 (1985) 340 [JETP Lett. 49 (1986)
419]; D. Friedan, Proc. Santa Barbara Workshop on Unified
String theories, eds. D. Gross and M. Green ( World Scientific,
Singapore, 1986).
\bibitem{fried}
D. Friedan, E.
Martinec and S. Schenker, Nucl. Phys. B 271 (1986) 93.
\bibitem{dan90}
G.S. Danilov, Phys.  Lett. B 257 (1991) 285;
Sov. J. Nucl.  Phys. 52 (1990) 727
[ Jadernaja Fizika 52 (1990) 1143 ];
Sov. J. Nucl. Phys. 49 (1989) 1106
[ Jadernaja Fizika 49 (1989) 1787].
\bibitem{danphr}
G.S. Danilov, Phys. Rev. D51 (1995) 4359 [Erratum-ibid. D52
(1995) 6201].
\bibitem{dan96}
G.S. Danilov, Phys. Atom. Nucl. 59 (1996) 1774 [Yadernaya
Fizika 59 (1996) 1837].
\bibitem{vec}
P. Di Vecchia, K. Hornfeck, M. Frau, A. Ledra and S. Sciuto,
Phys. Lett. B211 (1988) 301;
J.L. Petersen, J.R. Sidenius and A.K. Tollst{\'e}n,
Phys Lett. B 213 (1988) 30,
Nucl.  Phys.  B 317 (1989) 109.
\bibitem{dan93}
G.S. Danilov, JETP Lett. 58 (1993) 796 [Pis'ma JhETF 58 (1993)
790.]
\bibitem{danpr1}
G.S. Danilov, hep-th/0112022.
\bibitem{ver}
E. Verlinde and H. Verlinde, Phys. Lett. B192 (1987) 95;
Nucl. Phys B 288 (1987) 357.
\bibitem{as}
J. Atick, J. Rabin and A.  Sen, Nucl. Phys. B 299 (1988) 279;
J. Atick, G. Moore and A. Sen, Nucl. Phys. B 308 (1988) 1.
\bibitem{momor}
G. Moore and A. Morozov, Nucl. Phys. B 306 (1988) 387.
\bibitem{ber}
N. Berkovits, Nucl. Phys. B408 (1993) 43.
\bibitem{hok}
Eric D'Hoker and D.H. Phong, Phys. Lett. B529 (2002) 241; Nucl.
Phys. B 636 (2002) 3, 61; B 639 (2002) 129.
\bibitem{dangr}
G.S. Danilov, Phys. Lett. B 342 (1995) 73.
\bibitem{dannph}
G.S. Danilov, Nucl. Phys. B463 (1996) 443.
\bibitem{pol}
A.M. Polyakov,  Phys. Lett. B 103 (1981) 207, 210.
\bibitem{howe}
P.S. Howe, J. Phys. A 12 (1979) 393.
\bibitem{siegal}
C.L. Siegal, Topics in Complex Function Theory, Vol. 3 (New
York, Willey, 1973).



\end{thebibliography}
\end{document}